\documentclass[pra,aps,amsmath,amssymb,amsfonts,floatfix,groupeaddress,showpacs,twocolumn]{revtex4-1}

\usepackage{amsmath}
\usepackage{amssymb}
\usepackage{xspace}
\usepackage{graphicx}
\usepackage{grffile}
\usepackage{nicefrac}
\usepackage{color}

\graphicspath{{figs/}}

\newcommand{\bk}{\bf k}

\begin{document}

\title{Interface exchange processes in LaAlO$_3$/SrTiO$_3$
induced by oxygen vacancies}
\author{Malte Behrmann}
\affiliation{I. Institut f{\"u}r Theoretische Physik,
Universit{\"a}t Hamburg, D-20355 Hamburg, Germany}
\author{Frank Lechermann}
\affiliation{I. Institut f{\"u}r Theoretische Physik, Universit{\"a}t Hamburg, 
D-20355 Hamburg, Germany}

\begin{abstract}
Understanding the role of defects in oxide heterostructures is crucial for future
materials control and functionalization. We hence study the impact 
of oxygen vacancies (OVs) at variable concentrations on orbital- and spin exchange
in the LaAlO$_3$/SrTiO$_3$ interface by first principles many-body theory and
real-space model-Hamiltonian techniques. Intricate interplay between Hubbard $U$ and
Hund's coupling $J_{\rm H}$ for OV-induced correlated states is demonstrated. Orbital 
polarization towards an effective $e_g$ state with predominant local antiferromagnetic 
alignment on Ti 
sites near OVs is contrasted with $t_{2g}(xy)$ states with ferromagnetic 
tendencies in the defect-free regions. Different magnetic phases are identified, giving rise
to distinct net-moment behavior at low and high OV concentrations. This provides a
theoretical basis for prospective tailored magnetism by defect manipulation in oxide 
interfaces.
\end{abstract}


\maketitle

\section{Introduction}
There is strong evidence from recent experiments that oxygen vacancies play
a crucial role in the physics of heterostructures between LaAlO$_3$ (LAO) and 
SrTiO$_3$ (STO)~\cite{kal07,sie07,sal13,liu13,dav15}, as well as for 
surface~\cite{san11,mee11,mck14} and bulk~\cite{ric14} features of pure STO. For instance,
they may be relevant for ferromagnetic (FM) and superconducting order found in 
LAO/STO~\cite{rey07,bri07,li11,ari11,lee13}. OVs serve as electron 
dopants and can render an otherwise band-insulating environment metallic~\cite{li11}. 
The released charge from O$^{2-}$ fills the Ti$(3d)$ shell, which has $d^0$ occupancy
in stoichiometric STO. This introduces effects of electron correlation in strontium 
titanate~\cite{pen06,pav12,she12,lin13,lec14}, 
a compound adjacent to the Mott-insulating $3d^1$ $R$TiO$_3$ series (R: rare-earth ion) 
with a Ti$(3d^1)$ configuration. On similar grounds, the vacancy-induced doping enhances 
correlations in the LAO/STO interface. Puzzling interplay between itinerant and 
localized electrons in STO-based surfaces and interfaces is indeed suggested 
from scanning-tunneling spectroscopy~\cite{bre10,ris12,sit15}, magnetoresistance and
anomalous Hall-effect measurements~\cite{jos13}, resonant x-ray 
scattering~\cite{zho11,par13} and photoemission~\cite{san11,mee11,ber13}. 

A deeper comprehension of the defect influence on the interface phenomenology is motivated not just by basic research~\cite{sul14}. Since the physical properties of oxide 
heterostructures, ranging
from insulating and/or conducting to magnetic and/or superconducting, may be subtly tuned
by the presence of impurities, promising engineering aspects emerge.  Due to increasing control 
in detailed oxide-interface fabrication, new opportunities in high-response materials design
are within reach~\cite{bi14}. In view of future spintronics devices, selective magnetic 
activation on the nano scale by versatile ways of defect creation~\cite{bar12} may soon 
become available. 

Theoretical accounts of realistic LAO/STO heterostructures are challenging because of the unique
combination of complexity from the basic interacting quantum perspective and the structural 
bulk-to-interface setting. Calculations based on density functional theory (DFT) using hybrid 
functionals or employing static correlation effects from a Hartree-Fock-like treated Hubbard 
Hamiltonian ('+U') can reveal some relevant aspects of the intriguing electronic
structure~\cite{pen06,she07,pav12,she12,joh12,li13,jan14,lop15}. 
But there are two serious drawbacks to such extended Kohn-Sham schemes. Broken-symmetry 
states, i.e. long-range magnetic and/or charge orders, have to be stabilized often right 
from the start to address correlation effects. Second, several many-body hallmarks such 
as paramagnetic local-moment behavior, low-energy quasiparticle (QP) formation, 
band narrowing, and interplay between QPs and Hubbard bands are not incorporated due to a 
lack of frequency dependence in the local electronic self-energy $\Sigma$. 
Invoking DFT+dynamical mean-field theory (DMFT) overcomes these deficiencies, but, 
especially for defect environments treated by larger cells~\cite{lec14,gri14}, it
remains numerically expensive. It is important to note that many-body physics beyond standard
DFT-based approaches is here not only a detail, but essential for illuminating mechanisms of
future technological use. As a further relevant aspect, first-principles supercell computations 
to reveal defect influences are generally not perfectly suited to the problem at hand. They are 
restricted by the choice of the (often too small) cell size and can introduce artifacts because 
of the introduced short-range defect ordering. 

In this theoretical work we want to focus on the peculiar problem of OVs at the LAO/STO 
interface over a larger concentration range. Though a deeper relevance of these defects for the 
formation of the original quasi-twodimensional electron liquid is still under debate, several 
first-principles calculations have shown that even the stoichiometric heterostructure is metallic 
from partly filled Ti($t_{2g}$)-dominated bands at the Fermi level. Although there are other 
theoretical suggestions~\cite{mic12,li13,che13_2,ban13,ruh13,yu14}, OVs provide 
a natural way to explain the occurrence of interface 
ferromagnetism~\cite{she07,pav12,lec14,lop15}. In conventional DFT, an OV 
induces crystal-field lowered $e_g$-like impurity states 
on the neighboring Ti ions. Thinking intuitively, 
two limiting scenarios could apply depending on the concentration of vacancies in LAO/STO.
Lin and Demkov~\cite{lin13,lin14} studied the dilute-defect limit with only few oxygen 
defects, where $e_g$-like local moments on assumed Anderson/Kondo impurities may form. The 
latter can couple ferromagnetically via Ruderman-Kittel-Kasuya-Yosida (RKKY) interaction 
mediated by the itinerant $t_{2g}$ electrons. On the other hand, in a dense-defect 
limit, the physics is closer to a minimal two-orbital ($e_g$, $t_{2g}$) Hubbard 
model near quarter filling~\cite{lec14,pav13}. In a recent DFT+DMFT work~\cite{lec14} 
it was shown that in this limit, emerging FM order in the interface TiO$_2$ layer can 
indeed be explained by effective (Zener) double-exchange~\cite{zen51,and55} between an 
vacancy-induced $\tilde{e_g}$ orbital and an in-plane $t_{2g}(xy)$ orbital. 
Michaeli {\sl et al.}~\cite{mic12} also proposed Zener exchange between localized and 
itinerant states, but without referring to OVs as the source for localization.

Although experimentally the influence of OVs in STO-based materials is 
documented by monitoring physical response with varying oxygen partial 
pressure~\cite{ris12,liu13,mck14}, a good 
quantitative understanding of the OV concentration as well as definite information on the
location with respect to the interface~\cite{sie07,liu13} is still lacking.
To cope with the uncertainties in the number of OVs, we here perform 
investigations in a broad concentration range. In order to achieve this task, the correlated 
electronic structure is treated in 
a real-space framework allowing for in principle arbitrary vacancy configurations in 
number and arrangement. Depending on the concentration of OVs, we encounter different 
orbital- and spin-exchange regimes, that shed light on the emerging FM order at the 
LAO/STO interface. Local and non-local processes subject to a subtle interplay
between the Hubbard $U$ and Hund's $J_{\rm H}$ govern the OV-induced magnetism. Throughout
the paper all energies are given in electron volts.

The paper is organized as follows. To set the stage, we touch base with previous 
DFT+DMFT calculations~\cite{lec14} and start in Sec.~\ref{sec:jhund}  
with a brief view on the impact of the Hund's coupling $J_{\rm H}$ on the magnetic 
order in a n-type~\cite{hwa12} LAO/STO interface in the limit of high OV
concentration. In Sec.~\ref{sec:mod} the correlated real-space modeling to 
describe different defect concentrations is introduced. Results in the dilute
limit of a single oxygen vacancy as well as of two OVs in the interface are
presented in Sec.~\ref{sec:dilute}. Section~\ref{sec:more} deals with the evolution
of electron correlation and magnetism upon increasing the number of OVs 
from the dilute- to a dense-defect limit. The work closes with a discussion and summary in
Sec.~\ref{sec:sum}.

\section{Influence of  $J_{\rm H}$ in the dense-defect limit of oxygen vacancies in 
LAO/STO\label{sec:jhund}}
\begin{figure}[t]
\begin{center}
(a)\hspace*{-0.25cm}\includegraphics*[height=4cm]{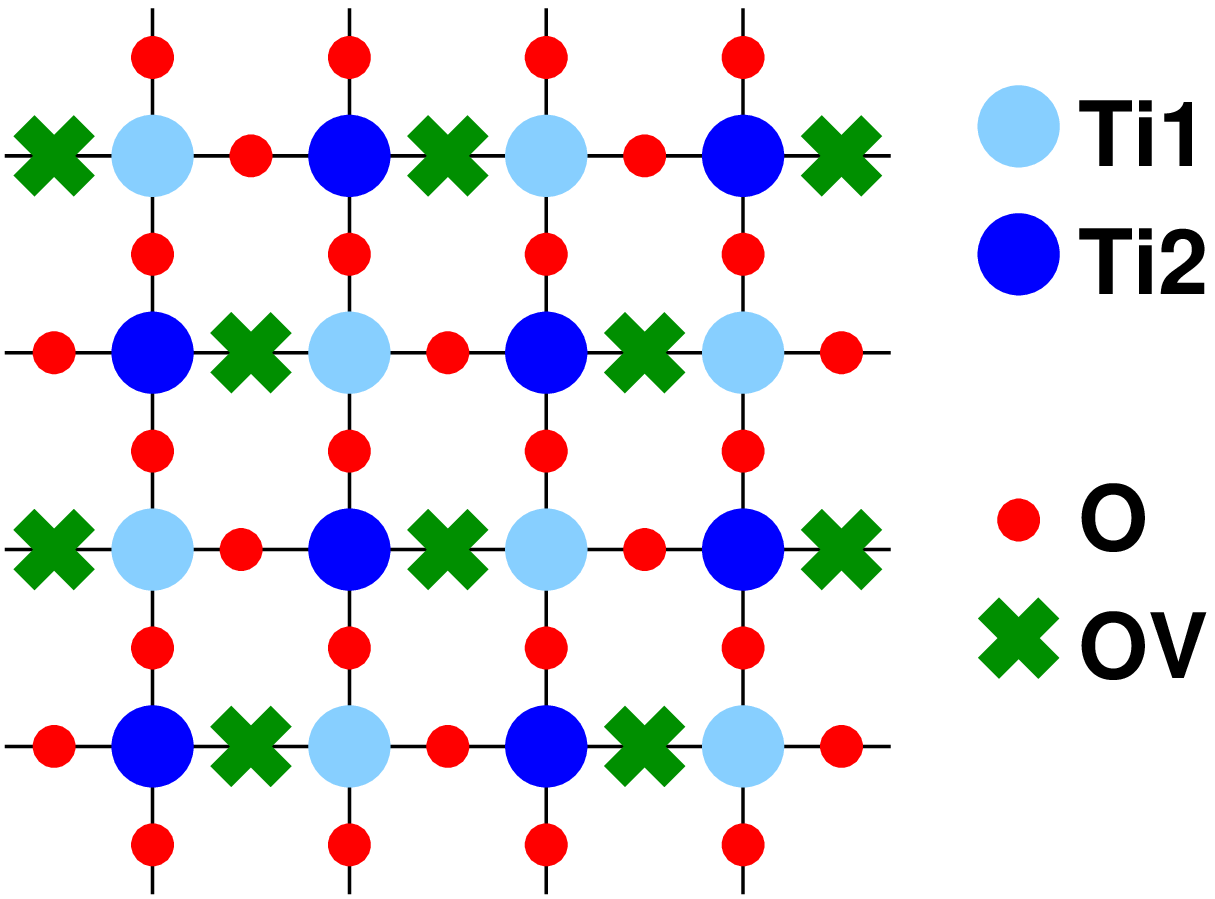}
(b)\hspace*{-0.25cm}\includegraphics*[height=4cm]{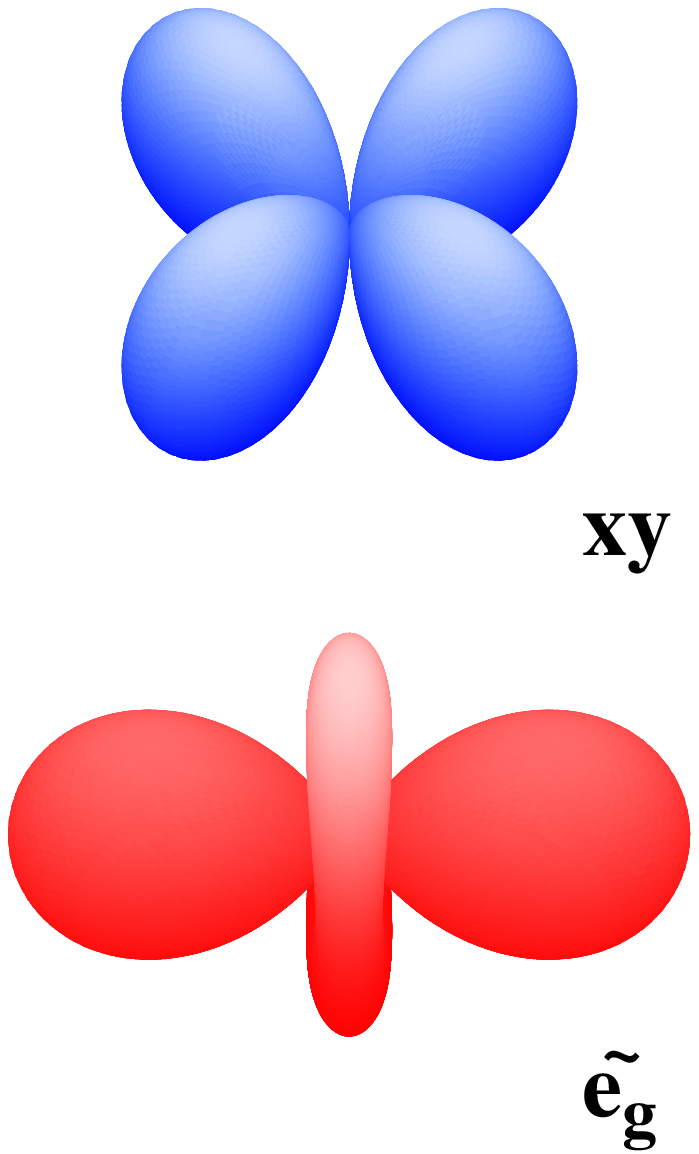}
\end{center}
\caption{(color online) (a) TiO$_2$ interface layer of n-type LAO/STO with 25\% oxygen 
vacancies utilized in supercell DFT+DMFT calculations~\cite{lec14}. Ions Ti1 and Ti2 
form the basis in the $\sqrt{2}$$\times$$\sqrt{2}$ primitive cell. (b) minimal relevant
Ti orbitals with 
$|\tilde{e_g}\rangle$$\,\sim\,$$0.55|z^2\rangle\pm 0.84|x^2$$-$$y^2\rangle$~\cite{lec14}.
\label{fig1:dense-defect}}
\end{figure}

A charge self-consistent DFT+DMFT study in a dense-defect limit of 25\% OVs in the 
TiO$_2$ interface layer was performed in Ref.~\onlinecite{lec14}. We define that limit by 
OVs exclusively located in the TiO$_2$ interface layer, with each Ti ion having one OV 
in bonding distance (see Fig.~\ref{fig1:dense-defect}). Here and throughout this paper Ti 
neighborhoods with more than one nearby OV are not considered. 

For the DFT part a mixed-basis pseudopotential 
framework is used and the DMFT impurity problems are solved by continuous-time quantum 
Monte Carlo (CT-QMC)~\cite{rub05,wer06,triqs_code,boe11}. A minimal correlated subspace 
was derived to consist of a two-orbital [$\tilde{e_g}$, $t_{2g}(xy)$] manifold located at
the interface Ti ions. Remaining $t_{2g}$ orbital degrees of freedom are included in 
more distant layers from the interface. The local Coulomb interactions in the 
interacting Hamiltonian with Slater-Kanamori parametrization, i.e., the Hubbard $U$ and Hund's 
exchange $J_{\rm H}$ were set to $U$=2.5 and $J_{\rm H}$=0.5, in line with other 
works~\cite{pav12}. 
A double-exchange-like (DE) mechanism is effective in stabilizing FM order within the interface.
Thereby the spin polarization is triggered by inter-orbital scattering between
the nearly-localized $\tilde{e_g}$ state and the less-localized $xy$ state. Because of this 
exchange mechanism, the fewer and more-itinerant $xy$ electrons exhibit stronger spin 
polarization. The self-consistently adapting Ti occupation results in a 
quarter filling ($n_{\rm Ti1, Ti2}$$\sim$1) of the two-orbital correlated subspace within
the interface layer. In total, each OV releases 2 electrons, which add to the 0.5 electrons 
per Ti ion due to the polar-catastrophe avoidance within LAO/STO heterostructures. 
In the supercell treatment of the dense-defect limit, which notably includes several 
TiO$_2$ layers, these two contributions adjust such that 1 electron per Ti settles in the 
interface TiO$_2$ layer. Hence the latter layer formally consists of Ti$^{3+}(3d^1)$ ions 
in the given defect limit. 

\begin{figure}[t]
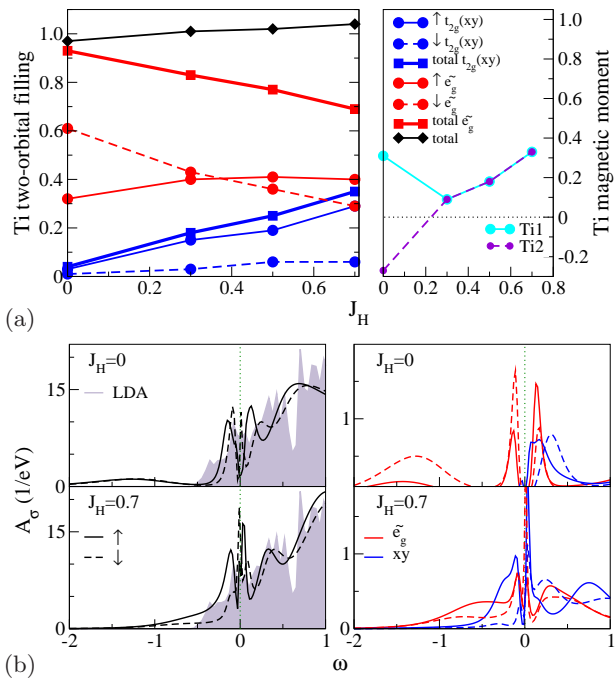

\begin{center}
(a)\hspace*{-0.3cm}\includegraphics*[width=8cm]{dmft-data.eps}\\[0.1cm]
(b)\hspace*{-0.3cm}\includegraphics*[width=8cm]{spectra.eps}
\end{center}
\caption{(color online) DFT-DMFT results depending on $J_{\rm H}$ for the dense-defect 
LAO/STO interface with magnetic order ($T$=180K). 
(a) Orbital occupations (left) and Ti magnetic moment (right). (b) Total (left) and local 
(right) spectral functions.\label{fig2:dmft}}
\end{figure}

Since the size of the Hund's coupling is key to a double-exchange-like mechanism for the 
onset of ferromagnetism, we here present results from varying $J_{\rm H}$, while keeping 
$U$=2.5 fixed. For more details on the calculational setup see Ref.~\onlinecite{lec14}. 
Usually the value of $J_{\rm H}$ is much less modified by screening processes than the 
Hubbard $U$ and therefore remains close to the magnitude in the free atom. Hence although for 
LAO/STO a Hund's coupling $J_{\rm H}$=0.5$-$0.6 is expected, we take the liberty of 
changing this value in order to assess it relevance for the underlying physics. 
Figure~\ref{fig2:dmft}(a) displays the orbital- and spin-dependent occupations with 
$J_{\rm H}$ in the magnetically ordered phase of the dense-defect limit. For 
$J_{\rm H}$$\gtrsim$0.4 ferromagnetism with local Ti moment 
$m$$\sim$0.2-0.4$\mu_{\rm B}$ occurs. But when $J_{\rm H}$ becomes smaller and 
eventually tends to zero, antiferromagnetic (AFM) order between the nearest-neighbor (NN) 
titanium ions sets in. In the latter regime the orbital polarization in favor of the
more localized $\tilde{e_g}$ level strongly increases towards nearly full polarization 
at $J_{\rm H}$=0.

Figure~\ref{fig2:dmft}(b) shows the $k$-integrated spectral function 
$A_\sigma(\omega)$=$\sum_{\bk}A_\sigma(\bk,\omega)$ in the limiting cases $J_{\rm H}$=0, 0.7 
for the whole supercell as well as for the correlated subspace of the 
Ti[$\tilde{e_g}$, $t_{2g}(xy)$] states. Concerning the correlation strength the Hund's coupling 
has a known model impact within a two-orbital system near quarter filling~\cite{lec07}. 
In the case of vanishing $J_{\rm H}$ stronger correlations occur, giving rise to a prominent 
lower Hubbard peak at $\sim$$-$1.3. This incoherent excitation is exclusively associated 
with the vacancy-induced $\tilde{e_g}$ state and resembles a similar feature in 
photoemission data~\cite{mee11,ber13,mck14}. Locally, hopping is nearly blocked on the 
interface Ti ions and residual metallicity is mainly provided by sites far from the 
interface. For rather large $J_{\rm H}$ the interface is well conductive, with a prominent 
QP peak at the Fermi level but an absent lower Hubbard peak. Coexistence of general 
metallicity with the latter incoherent excitation holds for reasonable intermediate values 
of $J_{\rm H}$~\cite{lec14}. The AFM order for $J_{\rm H}$=0 is completely carried by the 
$\tilde{e_g}$ orbital, while the FM order for $J_{\rm H}$=0.7 is dominantly carried
by the $xy$ orbital. These findings underline the importance of DE processes
in the formation of ferromagnetism at the LAO/STO interface in the dense-defect limit.

\section{Real-Space Approach to oxygen vacancies in LAO/STO\label{sec:rs}}
The scope is now broadened by addressing lower OV concentrations in the TiO$_2$ interface
layer. Performing large-scale incommensurate studies from first principles of the interplay between defects, 
disorder and correlations for multi-orbital lattice is 
nowadays still too expensive. We thus need a minimal setting that is geared to carry the key 
physics of the OV-doped LAO/STO interface. Therefore a model-Hamiltonian approach is used, 
inspired by the results of the DFT+DMFT investigation in the previous section.  

\subsection{Model and methodology\label{sec:mod}}
\begin{figure}[t]
\includegraphics*[width=8.25cm]{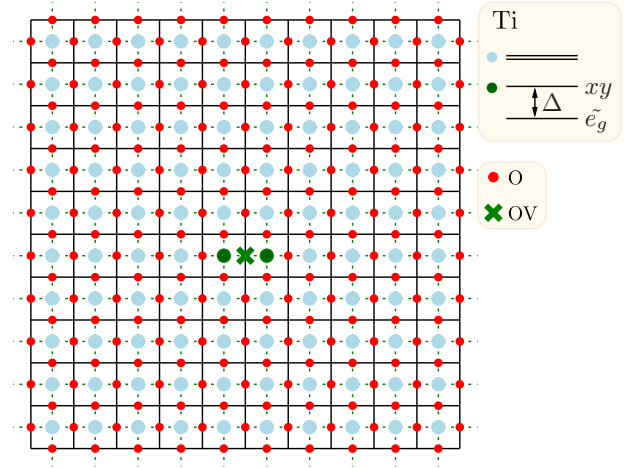}
\caption{(color online) Real-space two-orbital modeling on a 10$\times$10 TiO$_2$ square 
lattice for the n-type LAO/STO interface.
\label{fig3:setup}}
\end{figure}

A two-orbital Hubbard Hamiltonian based on the vacancy-induced effective $\tilde{e_g}$ 
state and the in-plane $t_{2g}(xy)$ state is employed on a 10$\times$10 square lattice with
$N_{\rm Ti}$=100 titanium ions, mimicking the interface TiO$_2$ layer 
(see Fig.~\ref{fig3:setup}). Only the Ti sublattice is treated explicitly and the oxygen degrees 
of freedom are integrated out within the chosen Hamiltonian form. Since the creation of
OVs amounts to electron doping, explicit involvment of remaining oxygen orbitals can be 
neglected to first approximation. Periodic boundary conditions are applied. Only intra-orbital 
NN hoppings are retained in the model. 

Lets focus first on the lattice scenario in the dense-defect limit, where each Ti site is 
affected by a nearby OV, to build up the model characteristics.
In line with Ref.~\onlinecite{lec14}, we choose $t_{\tilde{e_g}}$=$t_{xy}$=0.2 for the NN hoppings.
In contrast to a different modeling by Pavlenko {\sl et al.}~\cite{pav13}, our hopping amplitudes
from the projected-local-orbitals method~\cite{ama08} for higher OV concentrations are not 
strongly orbital dependent. From a noninteracting point of view, the crystal-field splitting
$\Delta$ between the $xy$ level and the vacancy-induced low-energy $\tilde{e_g}$ is the key
model parameter. Note that $\Delta$ is different from the usual octahedral crystal-field splitting
that is already vital in the stoichiometric compound. The latter energy splitting does not 
occur in the present defect model. Again from Ref.~\onlinecite{lec14}, we set $\Delta$=0.3.

Local electron-electron interactions via the Hubbard $U$ and Hund's coupling $J_{\rm H}$ are
applied at all Ti sites. The full Hamiltonian then reads
\begin{eqnarray}\label{eq:hubbardham}
H&=& -t \sum_{\langle i,j\rangle \alpha\sigma}c^\dagger_{i\alpha\sigma} 
c^{\hfill}_{j\alpha\sigma}
-\sum_{i\sigma}\Delta_i\left(n_{i,\beta,\sigma}-n_{i,xy,\sigma}\right)+\nonumber\\
&&\hspace*{-0.0cm}+U\,\sum_{i\alpha} n_{i\alpha\uparrow}n_{i\alpha\downarrow}+\nonumber\\
&&\hspace*{-0.0cm}+\frac 12 \sum \limits _{i,\alpha \ne \alpha',\sigma}
\Big\{U' \, n_{i\alpha \sigma} n_{i\alpha' \bar \sigma}
+ U'' \,n_{i\alpha \sigma}n_{i\alpha' \sigma}+\\
&&\hspace*{-0.0cm}+\left.J_{\rm H}\left(c^\dagger_{i\alpha \sigma} 
c^\dagger_{i\alpha' \bar\sigma}
c^{\hfill}_{i\alpha \bar \sigma} c^{\hfill}_{i\alpha' \sigma}
+c^\dagger_{i\alpha \sigma} c^\dagger_{i\alpha \bar \sigma}
 c^{\hfill}_{i\alpha' \bar \sigma} 
c^{\hfill}_{i\alpha' \sigma}\right)\right\}\;,\nonumber
\end{eqnarray}
where $c$$(c^\dagger)$ are creation (annihilation) operators, $n$=$c^\dagger c$,
$i,j$ are site indices, $\alpha,\alpha'$=$\beta$,$xy$ and 
$\sigma$=$\uparrow,\downarrow$ marks the spin projection, using
$U'$=$U-2J_{\rm H}$, $U''$=$U-3J_{\rm H}$. For the same Hubbard $U$, the strength of 
electronic correlations is usually weaker within slave-boson theory than within CT-QMC. If 
not otherwise stated the Hubbard $U$ is thus set to $U$=3 in all calculations. The Hund's 
coupling is set to $J_{\rm H}$=0.55, again with variations to smaller/larger values to trace 
its relevance. For the dense-defect limit, $\Delta_i$=$\Delta$ and $\beta$=$\tilde{e_g}$, i.e. 
the vacancy-induced $\tilde{e_g}$ crystal-field state is active on each Ti site.

In the latter limit every interface Ti ion has one neighboring OV and the given 
Hamiltonian is coherently applicable on each Ti site. However, when the defect number
is reduced, titanium sites without a nearby OV appear and at those Ti sites there
are no low-energy $\tilde{e_g}$ states. In the stoichiometric case the $e_g$ orbitals are 
strongly bound to O$(2p)$ and do not contribute either to states at the Fermi level or to any 
possible local-moment formation.
In order to keep the modeling simple, we make the following approximations when treating 
general defect cases in real space: (i) the Hamiltonian of form (\ref{eq:hubbardham}) is used 
throughout the lattice, (ii) the parametrization of $\Delta_i$ is
\begin{equation}
\Delta_i=\left\{\begin{array}{l@{:}l}
\Delta\;,\;\beta=\tilde{e_g}\quad &\quad\mbox{if an OV nearby}\\
     0\;\,\,,\;\beta=xz/yz  \quad &\quad\mbox{if no OV nearby} \end{array}\right.
\end{equation}
and (iii) multiple OVs around a Ti site are prohibited. 
The interpretation of (ii) is as follows: without a nearby OV, the Ti local low-energy 
electrons are mainly of $t_{2g}$ kind and thus the former $\tilde{e_g}$ degree of 
freedom takes over the role of an additional effective $t_{2g}$ orbital. This can
be justified by a notable hybridization between $\tilde{e_g}$ and $xz, yz$ in
the dense-defect case~\cite{lec14}. We neither change hoppings for Ti sites with or
without nearby OVs nor employ a concentration-dependent hopping modification. Such a more
detailed parametrization is hard to fix and the aim here is to work in a canonical 
two-orbital setting.

The final modeling step examines the concentration-dependent electron filling of 
a lattice with vacancy concentration $c$=$N_{\rm vac}/N_{\rm O}$, where $N_{\rm vac}$ 
is the number of OVs and $N_{\rm O}$=2$N_{\rm Ti}$ denotes the number of oxygen 
sites. Our electron count considers only the single TiO$_2$ 
interface layer, explicit charge fluctuations to more distant layers across the interface 
are neglected. In the dense-defect limit of the supercell treatment, DFT+DMFT yields
a filling of one electron at each Ti site in the interface TiO$_2$ layer. Supercell DFT 
calculations for the defect-free interface result in an itinerant two-dimensional electron 
system with a count of 0.5 electrons per interface Ti~\cite{lec14},  i.e. 50 electrons 
for the 100 Ti sites, in line with the polar-catastrophe avoidance. Putting these numbers 
together, we choose the linear-interpolation scheme 
$n_{\rm tot}$=$N_{\rm Ti}/2+N_{\rm vac}$ for the total lattice electron count $n_{\rm tot}$
at intermediate defect levels.

The interacting multi-orbital problem is solved by a real-space formulation of
the rotational-invariant slave-boson (RISB) mean-field 
method~\cite{li89,bue98,lec07,lec09,hua12}. It corresponds to single-site DMFT close to
zero temperature with a simpler impurity solver than the CT-QMC, allowing for local
self-energies $\Sigma$ with a linear frequency dependence and static terms. Renormalized QPs 
as well as local multiplets can be monitored in the interacting regime. Explicit intersite 
self-energy terms are neglected, but due to the coupling of all sites in the RISB 
self-consistency cycle effects of incoherency due to the distribution of defects are 
included. Our real-space approach is reminiscent of a single-orbital variant put into 
practice by Andrade {\sl et al.}~\cite{and09}. However instead of simplified 
Kotliar-Ruckenstein slave bosons~\cite{kot86} we here use the full rotational-invariant 
extension and elaborate on a multiorbital framework. Because of the model/method 
complexity no disorder averages are performed in this work. The calculations utilize
60 slave bosons and 17 lagrangian multipliers per site, 7700 variational parameters in total,
with dimension 400$\times$400 for the kinetic part of the Hamiltonian. 

In the following, the ordered magnetic moment $m$, the orbital moment $\tau$, the
paramagnetic local spin moment $m_{\rm PM}$ and the orbital polarization $\zeta$
are defined as
\begin{eqnarray}
m=\sum_\alpha m_\alpha=\sum_\alpha 
(\bar{n}_{\alpha\uparrow}-\bar{n}_{\alpha\downarrow})\;\;&,&\;
\tau=\sum_\sigma (\bar{n}_{\beta,\sigma}-\bar{n}_{xy,\sigma})\,\nonumber\\
m_{\rm PM}=\overline{\left(\sum_\alpha S_\alpha\right)^2}\;\;&,&\;
\zeta=
\frac{\sum_\sigma \bar{n}_{\beta,\sigma}}{\sum_\sigma \bar{n}_{xy,\sigma}}\quad,
\end{eqnarray}
where $S$ denotes the local spin operator and $\bar{\cal{O}}$=$\langle\cal{O}\rangle$.
Lattice-averaged values $Q_{\rm lat}$ of these quantities $Q$ are computed by 
$Q_{\rm lat}$=$Q/N_{\rm Ti}$.

\subsection{Dilute-defect limit\label{sec:dilute}}
Since our parametrization is adjusted to the DFT+DMFT results within the dense-defect limit, 
the model performance in the contrary limit of one or two OVs is of primary interest. For the 
case of two vacancies, a long-distance accommodation is chosen. Furthermore, they are placed 
on differently directed bars on the effective square lattice to minimize coherency effects.

\begin{figure}[b]
\includegraphics*[width=8cm]{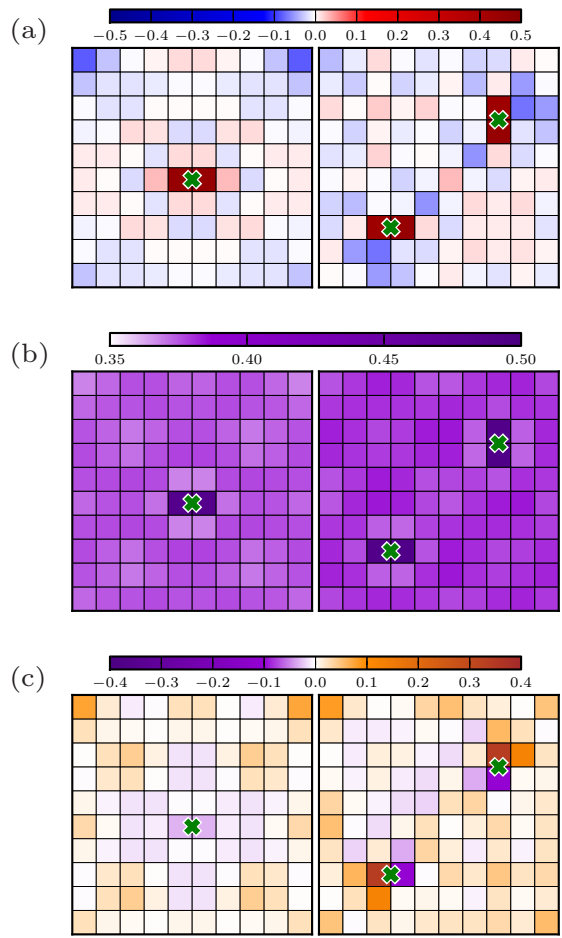}
\caption{(color online) Site-resolved quantities of interest for one OV (left) 
and two OVs (right) in the TiO$_2$ layer.
(a) PM orbital moment $\tau$, (b) PM local spin moment $m_{\rm PM}$, and (c) 
ordered magnetic moment $m$.
\label{fig4:fewdefect}}
\end{figure}

Figure~\ref{fig4:fewdefect} summarizes the obtained real-space results for key quantities 
in the PM phase as well as with magnetic order. Each small square resembles one Ti site and 
the oxygen sites are located in the middle of the enclosing bars. As expected, the orbital 
polarization towards the $\tilde{e_g}$ level for the Ti sites near vacancies is 
substantial. Away from the defect, especially in the two-OV case, there are, however, also 
oscillations in the orbital moment. The local PM spin moment is largest close to the 
OVs, understood from the well-localized correlated $\tilde{e_g}$ electrons, but attains a 
minimum just at second-nearest Ti sites. 

When initializing the RISB calculations with FM order, at saddle-point the solutions indeed 
reveal a small net FM lattice moment. But throughout the lattice the site-resolved ordered 
magnetic moment $m$ alternates. Interestingly, while $m$ for the single-OV case is still 
rather small at the defect, already the two-OV case displays a strong increase in the 
absolute value of the near-defect Ti magnetic moment. 
Furthermore, whereas for the single vacancy the spins on both NN Ti sites align
in the same direction, an AFM alignment emerges with two vacancies. This is surprising
since the nearby Ti sites in both scenarios show similar strong filling of the
localized $\tilde{e_g}$ state. Due to the strongly correlated regime, a significant
kinetic exchange $\sim$$t^2/U$ of AFM type is expected. Therefore the distance between OVs 
has to matter for the near-defect spin coupling. Structures with defects at closer range 
favor the local AFM alignment. The nonlocal exchange with the alternating $m$ throughout 
the lattice may also play a role in the local spin alignment. Thus there appears to be an 
intricate coupling between short- and longer-range exchange on the lattice.

Note that in general the overall system is rather susceptible to net magnetic order
even with only a small number of defects and small resulting moments. The electron count
per site reads $\bar{n}$=0.51(0.52) for one(two) OV(s), i.e. in principle the two-orbital
Hubbard modeling is close to one-eighth filling. Thus though the stoichiometric interface in 
DFT+DMFT does not show magnetic order~\cite{lec14}, it apparently only needs a low 
concentration of OVs to create the first magnetic instability in the correlated 
regime.
\begin{figure}[t]
\includegraphics*[width=8.5cm]{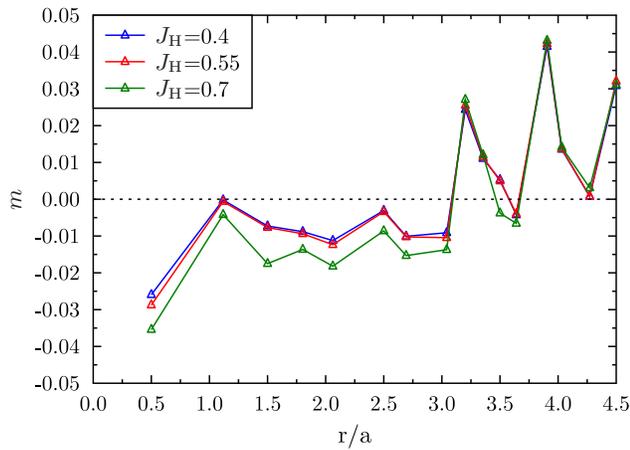}
\caption{(color online) Distance-dependent magnetic moment $m(r)$ with the oxygen vacancy 
at the origin in the magnetically ordered single-defect case for different choices 
of the Hund's coupling $J_{\rm H}$.\label{fig5:magosc}}
\end{figure}

In order to investigate the issue of magnetic exchange with distance in more detail,
the averaged ordered magnetic moment with spacing $r$ from the OV in the single-defect case
is plotted in Fig.~\ref{fig5:magosc}. The NN Ti site is located in distance 
$r_{\rm NN}$=0.5 in units of the lattice constant $a$. Though it is expected that results 
will dependent on the chosen lattice size, there are significant variations within the 
shorter- and longer-range regions. For $d$$\sim$2.5 the moments predominantly change their 
sign and the spin coupling switches from FM- to AFM-like. Let's assume here a 
Fermi-wave-vector $k_F$ modulated RKKY exchange with $J(d)$$\sim$$\cos(2k_F\,d)$ in the 
low-density limit for the sea of conduction electrons~\cite{mic12}. This would correspond to 
$k_F$$\sim$$\pi/10$ in reciprocal units, more or less commensurable with the noted one-eighth 
filling. But one has to keep in mind that the local ordered moment in our single-OV case 
is too small for a serious application of the standard RKKY picture. Still, for the record,
there are indications of RKKY-like exchange taking place in the dilute-defect limit.
Concerning the influence of Hund's coupling, not surprisingly, a larger $J_{\rm H}$ increases
the spin moments since it tends to locally align the contributions from the orbital degrees of freedom.

\subsection{Electron correlation and magnetism from the dilute- to the dense-defect limit\label{sec:more}}
We now investigate the regime between the dilute-defect limit ($c$=0.005, 0.01)
and the dense-defect limit ($c$=0.25). Additional configurations with randomly 
distributed OVs for intermediate concentrations are constructed, prohibiting local Ti 
neighborhoods with more than one vacancy in NN distance, respectively. 
\begin{figure}[b]
\includegraphics*[width=8.5cm]{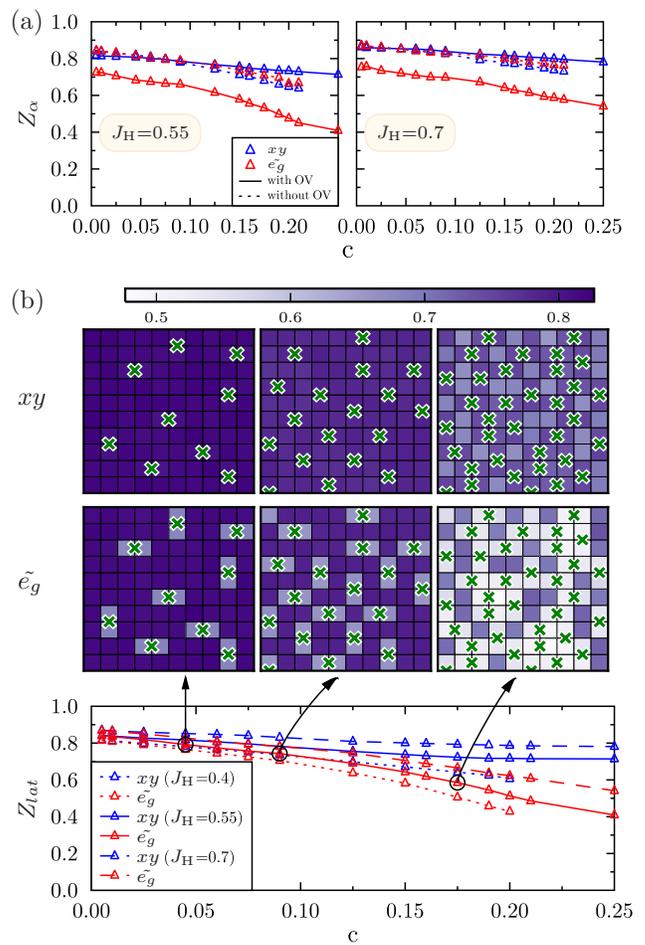}
\caption{(color online) Orbital-resolved QP weight $Z$ vs concentration of oxygen 
vacancies (OVs) in the PM phase. Total electron count increases from $n_{\rm tot}$=51 (1 OV, 
$c$=0.005) to $n_{\rm tot}$=100 (50 OVs, $c$=0.25). (a) Average $Z_\alpha$ for states on 
Ti sites with and without nearby OVs for $J_{\rm H}$=0.55 (left) and $J_{\rm H}$=0.7 (right).
(b) Site-resolved $Z$ for selected dopings and $J_{\rm H}$=0.55 (top) as
well as lattice $Z_{\rm lat}$ for different choices of $J_{\rm H}$ (bottom).
\label{fig6:qpweight}}
\end{figure}
In the following, site-resolved and site-averaged data is presented. It proves
instructive not only to perform full lattice averages, but also to differentiate between
the two groups of Ti sites, i.e., those with and those without nearby OVs.

Figure~\ref{fig6:qpweight} provides a measure of the correlation strength by displaying
the paramagnetic QP weight 
$Z$=$\left[1-\frac{\partial}{\partial\omega}\Sigma\,\right]_{\omega=0}^{-1}$. 
Because of the low electron count at low OV concentrations the lattice QP weight starts off 
with values close to the noninteracting limit $Z_{\rm lat}$=1. With more vacancies 
and increased electron doping, general electronic correlations become stronger down to 
$Z_{\rm lat}$$\sim$0.4 in the dense-defect case. For the considered electron-doping regimes 
the Hund's coupling $J_{\rm H}$ on average weakens correlations, in line with DFT+DMFT 
results from Sec.~\ref{sec:jhund}. For any doping, orbital- and/or site-selective 
Mott transitions remain absent. Not surprisingly, electrons in the $\tilde{e_g}$ orbitals 
are nonetheless stronger correlated since they are more localized due to the lowered crystal field.
While for the $xy$ state an obvious site discrimination occurs only at higher doping, the
$\tilde{e_g}$ electrons near OVs are already much heavier at the lowest doping. 
Interestingly the $xy$ electrons eventually become stronger correlated {\sl away}
from near-OV titanium. Thus the vacancy influence on correlations is twofold:
it locally strengthens the $\tilde{e_g}$ correlation and nonlocally fosters the 
$xy$ correlations.

\begin{figure}[t]
\hspace*{-0.4cm}\includegraphics*[width=8.75cm]{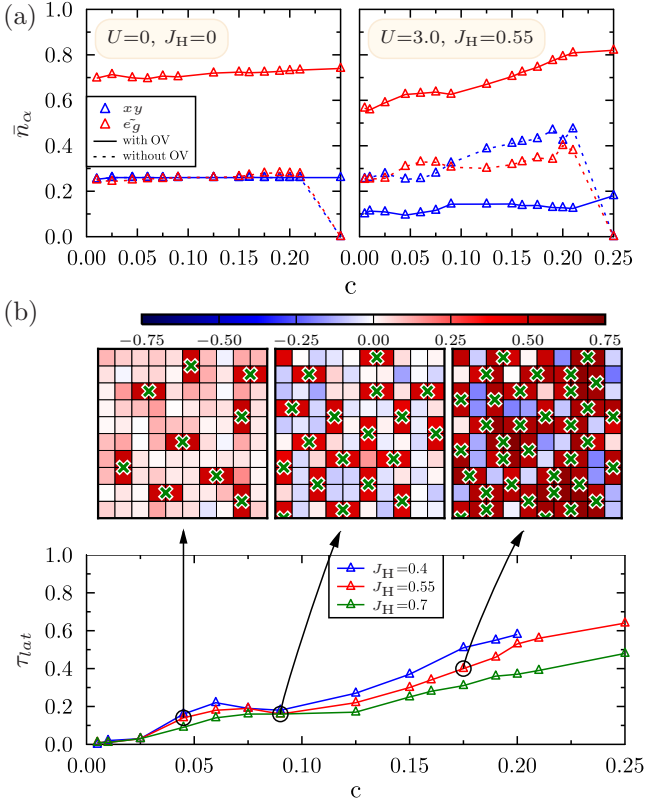}
\caption{(color online) Orbital-resolved occupation in the PM phase.
(a) Average occupations of Ti sites with and without nearby OVs for the
noninteracting (left) and interacting (right) case. 
(b) Concentration-dependent orbital moment $\tau$ in real space (top) and lattice-averaged 
(bottom).
\label{fig7:orb}}
\end{figure}

As mentioned, the site-dependent correlation strength is of course related to
the local orbital filling, shown in Fig.~\ref{fig7:orb} from low to higher OV-induced
electron doping. For comparison, orbital occupations in the noninteracting case
($U$=$J_{\rm H}$=0) are additionally depicted. For the Ti sites with nearby OVs a
strong orbital polarization towards $\tilde{e_g}$ already at low dopings is obvious,
even without local Coulomb interactions. Rather 
independent of the number of vacancies, a polarization $\zeta$$\sim$2.8 holds for these
sites in the noninteracting problem. With interactions this orbital polarization is 
substantially increased for all doping levels because of the crystal-field renormalization
via the electronic self-energy. It still gradually decreases from $\zeta$$\sim$6 in the 
dilute limit to $\zeta$$\sim$4 in the dense limit. 
\begin{figure}[t]
\hspace*{-0.4cm}\includegraphics*[width=8.75cm]{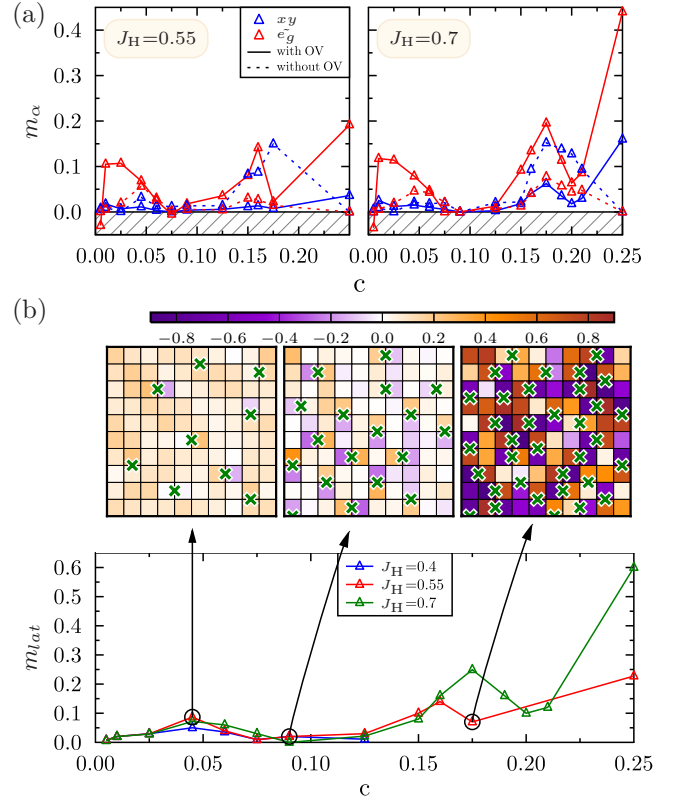}
\caption{(color online) Orbital-resolved magnetic moment $m$. 
(a) Average magnetic moment of Ti sites with and without nearby OVs for $J_{\rm H}$=0.55
(left) and $J_{\rm H}$=0.7 (right). (b) Concentration-dependent magnetic moment $m$ in 
real space (top) and lattice-averaged (bottom).
\label{fig8:mag}}
\end{figure}

Due to the absence of local crystal fields a subtle competition between both orbital 
degrees of freedom occurs at the remaining Ti sites. However, remember that in this region the 
'$\tilde{e_g}$' orbital inherits the role of an additional $t_{2g}$ orbital in
our modeling. The noninteracting case does not reveal any finite orbital moment for any 
OV concentration. By including Coulomb interactions, low doping from the dilute limit 
slightly disfavors the $xy$ orbital. But interestingly, at $c_{\rm p}$$\sim$0.08
the $xy$ orbital takes over the lead in occupation. Thus also here a nonlocal impact of 
the OV-induced correlations shows up, breaking the orbital degeneracy at Ti sites away 
from the defects. Since in the dense-defect limit every Ti ion has one nearby OV, the class
of defect-free Ti sites disappears and its nominal filling then, of course, vanishes. 
Intuitively, the doping-dependent change of orbital-filling hierarchy for this Ti class  
can be understood as follows. At very low dopings, the interacting system tries to 
put more electrons into the $\tilde{e_g}$ level, since there they can occasionally enjoy the lower crystal field close to an OV. Yet at some concentration with increased
electron filling, it is more beneficial to put electrons which like to visit  
defect-free regions in the overall less occupied $xy$ levels to minimize the Coulomb 
interaction and gain kinetic energy.

The real-space variation of the site-dependent orbital moment $\tau$ underlines these 
findings [see Fig.~\ref{fig7:orb}(b)]. Beyond the critical doping level $c_{\rm p}$
there is a qualitative change in the polarization of the 'interstitial' region towards
$xy$. A shoulder in the lattice-averaged orbital moment $\tau_{\rm lat}$ is located around
$c_{\rm p}$. Of course, every increase in OVs, with strong orbital polarization 
towards $\tilde{e_g}$ nearby, renders $\tau_{\rm lat}$ further monotonically growing with 
$c$. 
Give or take, the averaged influence of $J_{\rm H}$ on the orbital moment is as expected 
from multiorbital Hubbard models, i.e. it works against the crystal field and tries to 
wash out orbital polarization~\cite{lecproc}.

The orbital- and site-resolved magnetic moment exhibits an even more intricate structure
with respect to the vacancy concentration (cf. Fig.~\ref{fig8:mag}). Starting from the 
dilute-defect limit the Ti sites with nearby OVs can be divided into two subclasses. One spin thereof 
compensates their net moment by AFM alignment, the other one displays an FM alignment
with smaller moments. Therefrom a still sizable net FM moment near $c$$\sim$0.05 results,
whereby the RKKY-like exchange noted in Sec.~\ref{sec:dilute} can be held responsible.
Intriguingly, close to $c_{\rm p}$ this net FM moment vanishes, accompanied by the
disappearance of the Ti-site subclass with FM alignment near OVs. In addition, the nearly 
exclusive AFM alignment at OVs above $c_{\rm p}$ comes along with zero spin polarization
on the remaining Ti sites in the concentration range 0.08$<$$c$$<$0.13. Thus the RKKY-like
driven FM phase is followed by a phase region of separated AFM pairs with zero net moment.
\begin{figure}[t]
\begin{center}
\includegraphics*[width=8.5cm]{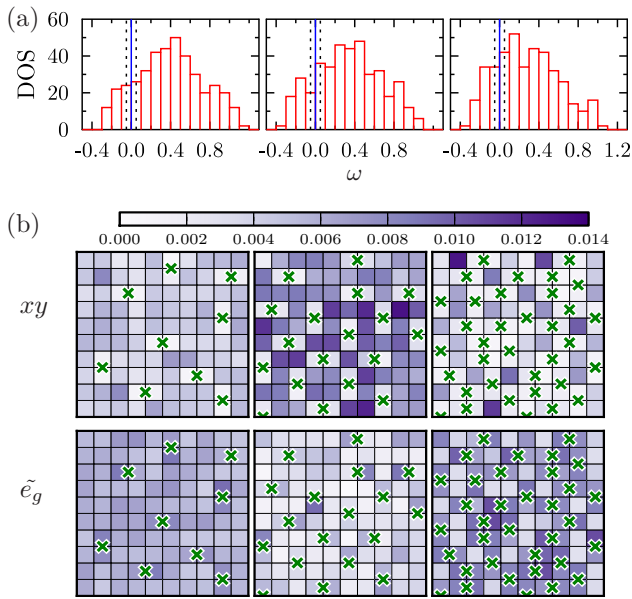}
\end{center}
\caption{(color online) Low-energy PM spectral information for dopings
$c$=(0.045, 0.09, 0.175) (left to right) using $J_{\rm H}$=0.55.
Orbital-resolved site-contribution $|C_{im}|^2$ to the renormalized QP 
state $|\Psi_{\rm QP}\rangle$=$\sum_{im}C_{im}|\phi_{im}\rangle$ in
an energy window $[-0.05,0.05]$ around the Fermi level. 
\label{fig9:fatbands}}
\end{figure}
Though, as discussed before, the $xy$ {\sl orbital} polarization in the 'interstitial' is 
already active, in this concentration range there is no novel exchange mechanism yet strong 
enough to {\sl spin} polarize the Ti sites aways from the defects. This changes for 
$c_{\rm DE}$$\gtrsim$0.13, when eventually the latter sites build up a considerable 
$xy$-dominated magnetic moment. Effective non-local double-exchange between the 
more-localized $\tilde{e_g}$ electrons near OVs and the more-itinerant $xy$ electrons away 
from OVs yields net ferromagnetism. Also, part of the AFM alignment of the near-defect
Ti sites breaks up and these sites join in the contribution to the FM order. Interestingly,
there is a minimum in the averaged $\tilde{e_g}$ magnetic moment on the way to the dense-defect
limit. This may be explained by the competition between FM double-exchange and AFM kinetic
exchange on Ti near OVs. The increasing electron filling re-strengthens the kinetic exchange
close to the defect in the now FM-polarized environment, i.e. a re-formation of AFM pairs
occurs. A larger Hund's coupling shifts that minimum to higher electron dopings
[see Fig.~\ref{fig8:mag}(a)], i.e. the local $\tilde{e_g}$ occupation has to be closer to the 
kinetic-exchange favored half-filling [cf. Fig.~\ref{fig7:orb}(a)] to overcome the 
$J_{\rm H}$-supported DE processes.
We also checked the generic influence of a smaller $U$=2.5 and encountered an overall somewhat 
reduced magnetic moment and a weakening of the $xy$ magnetism based on the nonlocal double-exchange.

For the FM phases, our revealed lattice magnetic moment 
$m_{\rm lat}$$\sim$0.1-0.2$\mu_{\rm B}$ is in very good agreement with experimental 
findings~\cite{lee13}. So for moderate $J_{\rm H}$ the nonlocal polarization effect of OVs 
fosters a sizable $xy$-dominated magnetic moment on defect-free Ti sites above a
concentration $c_{\rm DE}$$\sim$0.13. This nonlocal double-exchange process extends the 
local DE mechanism from coherent systems, here active in the dense-defect limit. 
Note that the obtained magnetic moment $m$$\sim$0.2$\mu_{\rm B}$
for $J_{\rm H}$=0.55 in the latter limit is in excellent agreement with the former 
DFT+DMFT results (cf. Fig.~\ref{fig2:dmft}), highlighting the consistency of the model.
However, there is a difference in the orbital contributions, since in the real-space RISB
model the $\tilde{e_g}$ level is more strongly spin-polarized than $xy$. This may be explained by
the fact that fluctuations and their correlations, relevant for assessing the DE processes in
detail, are underestimated in simplified RISB compared to DMFT with a CT-QMC solver. It 
could also be that scattering in additional TiO$_2$ layers, which is not included in the
real-space modeling, supports the $xy$ spin polarization.

Finally, we address spectral features at low energy since, e.g., the question arises about the
different site and orbital contributions to the resulting metallicity. For selective
dopings, Fig.~\ref{fig9:fatbands} shows the total QP density of states (DOS) as well as 
the site- and orbital-resolved QP spectral weight within a small energy window around the 
Fermi level, both in the PM regime. The real-space/orbital resolution is naturally derived
from analyzing the low-energy lattice eigenvectors of the renormalized kinetic Hamiltonian.

At the concentration just above $c_{\rm p}$ the total DOS displays a pseudogap-like feature 
at the Fermi level. From a Stoner argumentation, this reduced spectral weight at 
$\varepsilon_F$ is in line with a vanishing of ferromagnetism in this
concentration regime. In the case of higher OV numbers, i.e. higher electron dopings, the 
low-energy density of states rises again within the DE-FM region. As generally expected the 
$xy$ weight is mainly located in the defect-free regions, and the $\tilde{e_g}$ weight stems 
dominantly from defect-near regions. 
Close to the dilute-defect limit, the overall $xy$ low-energy weight is lower, but it
becomes dominant just above $c_{\rm p}$. There the $\tilde{e_g}$ electrons appear most 
localized, giving eventually rise to the pair-AFM phase. For $c_{\rm DE}$$\gtrsim$0.13 
eventually the $\tilde{e_g}$ low-energy contribution again overcomes the $xy$ one, marking 
the intriguing scattering regime of the effective double-exchange region.

\section{Summary and Discussion\label{sec:sum}}
This work examined the key effects of the electronic structure reconstruction in the
LAO/STO interface due to the presence of OVs. Different orbital and spin
exchange processes are identified for varying OV concentrations. From the revealed 
and expected magnitudes of hoppings, crystal fields and Coulomb interactions a 
straightforward picture of fully localized (Kondo-like) electrons near OVs is not evident. 
Charge self-consistent DFT+DMFT for LAO/STO supercells in a dense-defect limit yields a 
dichotomy between OV-induced heavier $\tilde{e_g}$ states as well as $xy$ states with a 
higher QP weight. But itinerancy remains a common feature when including a finite Hund's
coupling $J_{\rm H}$. The latter not only triggers the correlation strength but fosters 
ferromagnetism above a threshold doping through effective nonlocal and local 
double-exchange processes. Additionally, the strongly correlated dense-defect limit is 
spectrally marked by a lower Hubbard peak of $\tilde{e_g}$ kind.
For generic OV numbers in a 10$\times$10 TiO$_2$ model interface our correlated real-space 
RISB approach elucidates further mechanisms due to the system separation into Ti sites near 
and away from OVs. Coulomb interactions are relevant to trigger intricate 
(nonlocal) orbital polarization processes, especially at the defect-free Ti sites. 

\begin{figure}[t]
\begin{center}
\includegraphics*[width=8.75cm]{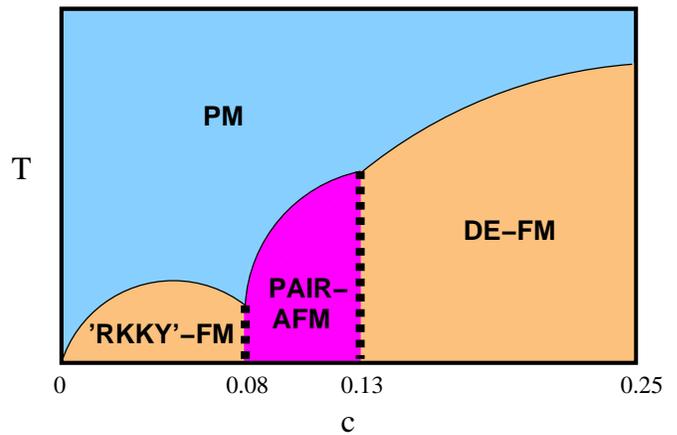}
\end{center}
\caption{(color online) Rough sketch of a finite-temperature ($T$) magnetic phase diagram
for OV concentrations $c$ based on the real-space RISB modeling of the LAO/STO interface.
Dashed phase boundaries of the pair-AFM phase mark the possible occurrence of
coexistence regions with the nearby FM phases. 
\label{fig10:phases}}
\end{figure}

Concerning magnetic order, the schematic phase diagram in Fig.~\ref{fig10:phases} summarizes
the main findings. Already in the dilute limit of very few vacancies, oscillations in the 
sign of the magnetic moments with distance from the OV can be detected. Near OVs two 
subclasses of Ti sites appear, one favors local AFM and the other local FM alignment. An 
effective RKKY-like exchange mechanism weakly spin polarizes the system, giving rise to a 
finite FM net moment. We note that the present RKKY ordering
is not conventional in the sense that the involved local moments are comparatively small.
Due to the delicate exchange mechanism, the expected Curie temperature $T_c$ associated with 
this phase is rather low. Above an interface vacancy concentration $c_p$$\sim$0.08, the 
pairs of AFM-aligned Ti sites dominate the lattice and the spin polarization in the regions 
without OVs disappears until at $c_{\rm DE}$$\sim$0.13 the double-exchange becomes strong 
enough to polarize the 'interstitial', now with dominant $xy$ character. In addition, the
DE mechanism is effective in switching local AFM pairs to FM alignment. A more robust 
ferromagnetic order sets in, with a supposedly much larger $T_c$, and continues to be stable 
up to the dense-defect limit. The value of $J_{\rm H}$ influences the competition between
$\tilde{e_g}$ filling-controlled re-strengthened AFM-like kinetic exchange and the DE
processes near OVs in this novel DE-FM phase. We did not delve into the possible phase 
transitions among the three phases. Because of the overall defect system without
a coherent local order parameter and straightforward symmetry distinction, first-order(-like) 
transitions with coexistence regions are expected (cf. Fig.~\ref{fig10:phases}).

Our obtained behavior with increasing OV concentration shares several features with
experimental results. It is generally in accordance with the found key dependence of magnetism
on electron doping in STO-based materials. An interplay of AFM and FM tendencies has
been recently identified by Bi {\sl et al.}~\cite{bi14}. The different experimental results 
concerning the range of stability for LAO/STO ferromagnetism may be related to substantial 
differences in the number of vacancies in the respective samples. Whereas nearly 
stoichiometric interfaces are susceptible to the low-$T_c$ RKKY-like FM 
phase~\cite{fit11,ron14}, 
OV-rich samples can stabilize the DE-FM phase with the surprisingly high $T_c$ near room 
temperature~\cite{ari11,bi14}. Of course, such different ferromagnetism may also emerge
in very inhomogeneous samples. A theoretical resolution of phase separation on a 
larger lattice scale based on the present modeling is, however, numerically hard to achieve.
Concerning the crucial concentration dependence, unique behavior connected to critical
electron densities has been revealed in magnetotransport measurements~\cite{jos13}. In
that respect it would be very interesting to trace in detail the ferromagnetism in applied
magnetic field $B$ within a group of samples with different OV concentrations, or to perform in-situ 
monitoring with oxygen pressure. For instance, the pair-AFM phase could be transformed to
FM order by a larger field $B$. 

We have shown that itinerancy, polarized orbital degrees of freedom and magnetic order 
naturally go together in LAO/STO interfaces with OVs. The realistic many-body
physics remains challenging and needs further work. From theory, the treatment of the 
full Ti$(3d)$ shell for the demanding supercell- and real-space computations would allow
for a more detailed orbital resolution. Including the impact of spin-orbit coupling and
the competition of the revealed phases with superconductivity is a natural
further modeling step. In addition, with even larger numerical effort by further extension 
to a cluster-RISB framework~\cite{lec07}, one could introduce a two-site cluster in real space
for each pair of Ti sites linked by an OV. Then possible singlet formation, contrary to the here-described 
local pair-AFM state, via intersite self-energies would be describable. Note that we 
have limited our work to single-OV defects. An investigation of multi-OV configurations up to 
vacancy-clustered regions on very large real-space lattices will surely be appreciated future work. 
In this respect, experimental information on defect concentrations, locations, and 
arrangements beyond currently available data is greatly needed. 

Generally, controlling the defect structure within interfaces of oxide heterostructures will be 
pivotal to eventually engineering technological applications with designated response behavior.
Recent findings of room-temperature ferromagnetism and enhanced photocatalytic performance in 
Ti-defected TiO$_2$ Anatase~\cite{wan15} are a further example for the relevance and potential 
of defect-induced oxide physics beyond the high-$T_c$ cuprate paradigm. The here-documented 
sensitivity of the (magnetic) electronic 
structure to the OV concentration could be useful not just for designing LAO/STO-based magnetic order 
at elevated temperatures. Charge and magnetic writing, flexible spintronic switches, sensor 
technology, and possible multiferroic response are only a few further optional engineering 
directions. The idea of creating atom-resolved orbital and spin polarization by controlled 
defect manipulation within a well-defined interface region has so far not been translated into 
practicable device physics. Defect control of emerging interface phases could be complementary 
to the technological potential of adatom-driven surface phenomena. The real-space
approach presented here is especially suited to simulate and direct such design and control of challenging 
correlated materials on a nano scale.

\begin{acknowledgments}
We are grateful to T. Kopp, N. Pavlenko and C. Piefke for helpful discussions. 
This research was supported by the Deutsche Forschungsgemeinschaft through FOR1346 and 
SFB925. Computations were performed at Regionales Rechenzentrum (RRZ) of the University of Hamburg and at the JUROPA 
Cluster of the J\"ulich Supercomputing Centre (JSC) under Project No. hhh08.
\end{acknowledgments}
\bibliographystyle{apsrev4-1}
\bibliography{bibextra}

\begin{thebibliography}{61}%
\makeatletter
\providecommand \@ifxundefined [1]{%
 \@ifx{#1\undefined}
}%
\providecommand \@ifnum [1]{%
 \ifnum #1\expandafter \@firstoftwo
 \else \expandafter \@secondoftwo
 \fi
}%
\providecommand \@ifx [1]{%
 \ifx #1\expandafter \@firstoftwo
 \else \expandafter \@secondoftwo
 \fi
}%
\providecommand \natexlab [1]{#1}%
\providecommand \enquote  [1]{``#1''}%
\providecommand \bibnamefont  [1]{#1}%
\providecommand \bibfnamefont [1]{#1}%
\providecommand \citenamefont [1]{#1}%
\providecommand \href@noop [0]{\@secondoftwo}%
\providecommand \href [0]{\begingroup \@sanitize@url \@href}%
\providecommand \@href[1]{\@@startlink{#1}\@@href}%
\providecommand \@@href[1]{\endgroup#1\@@endlink}%
\providecommand \@sanitize@url [0]{\catcode `\\12\catcode `\$12\catcode
  `\&12\catcode `\#12\catcode `\^12\catcode `\_12\catcode `\%12\relax}%
\providecommand \@@startlink[1]{}%
\providecommand \@@endlink[0]{}%
\providecommand \url  [0]{\begingroup\@sanitize@url \@url }%
\providecommand \@url [1]{\endgroup\@href {#1}{\urlprefix }}%
\providecommand \urlprefix  [0]{URL }%
\providecommand \Eprint [0]{\href }%
\providecommand \doibase [0]{http://dx.doi.org/}%
\providecommand \selectlanguage [0]{\@gobble}%
\providecommand \bibinfo  [0]{\@secondoftwo}%
\providecommand \bibfield  [0]{\@secondoftwo}%
\providecommand \translation [1]{[#1]}%
\providecommand \BibitemOpen [0]{}%
\providecommand \bibitemStop [0]{}%
\providecommand \bibitemNoStop [0]{.\EOS\space}%
\providecommand \EOS [0]{\spacefactor3000\relax}%
\providecommand \BibitemShut  [1]{\csname bibitem#1\endcsname}%
\let\auto@bib@innerbib\@empty
\bibitem [{\citenamefont {Kalabukhov}\ \emph {et~al.}(2007)\citenamefont
  {Kalabukhov}, \citenamefont {Gunnarsson}, \citenamefont {B{\"o}rjesson},
  \citenamefont {Olsson}, \citenamefont {Claeson},\ and\ \citenamefont
  {Winkler}}]{kal07}%
  \BibitemOpen
  \bibfield  {author} {\bibinfo {author} {\bibfnamefont {A.}~\bibnamefont
  {Kalabukhov}}, \bibinfo {author} {\bibfnamefont {R.}~\bibnamefont
  {Gunnarsson}}, \bibinfo {author} {\bibfnamefont {J.}~\bibnamefont
  {B{\"o}rjesson}}, \bibinfo {author} {\bibfnamefont {E.}~\bibnamefont
  {Olsson}}, \bibinfo {author} {\bibfnamefont {T.}~\bibnamefont {Claeson}}, \
  and\ \bibinfo {author} {\bibfnamefont {D.}~\bibnamefont {Winkler}},\
  }\href@noop {} {\bibfield  {journal} {\bibinfo  {journal} {Phys. Rev. B}\
  }\textbf {\bibinfo {volume} {75}},\ \bibinfo {pages} {121404(R)} (\bibinfo
  {year} {2007})}\BibitemShut {NoStop}%
\bibitem [{\citenamefont {Siemons}\ \emph {et~al.}(2007)\citenamefont
  {Siemons}, \citenamefont {Koster}, \citenamefont {Yamamoto}, \citenamefont
  {Harrison}, \citenamefont {Lucovsky}, \citenamefont {Geballe}, \citenamefont
  {Blank},\ and\ \citenamefont {Beasley}}]{sie07}%
  \BibitemOpen
  \bibfield  {author} {\bibinfo {author} {\bibfnamefont {W.}~\bibnamefont
  {Siemons}}, \bibinfo {author} {\bibfnamefont {G.}~\bibnamefont {Koster}},
  \bibinfo {author} {\bibfnamefont {H.}~\bibnamefont {Yamamoto}}, \bibinfo
  {author} {\bibfnamefont {W.~A.}\ \bibnamefont {Harrison}}, \bibinfo {author}
  {\bibfnamefont {G.}~\bibnamefont {Lucovsky}}, \bibinfo {author}
  {\bibfnamefont {T.~H.}\ \bibnamefont {Geballe}}, \bibinfo {author}
  {\bibfnamefont {D.~H.~A.}\ \bibnamefont {Blank}}, \ and\ \bibinfo {author}
  {\bibfnamefont {M.~R.}\ \bibnamefont {Beasley}},\ }\href@noop {} {\bibfield
  {journal} {\bibinfo  {journal} {Phys. Rev. Lett.}\ }\textbf {\bibinfo
  {volume} {98}},\ \bibinfo {pages} {196802} (\bibinfo {year}
  {2007})}\BibitemShut {NoStop}%
\bibitem [{\citenamefont {Salluzzo}\ \emph {et~al.}(2013)\citenamefont
  {Salluzzo}, \citenamefont {Gariglio}, \citenamefont {Stornaiuolo},
  \citenamefont {Sessi}, \citenamefont {Rusponi}, \citenamefont {Piamonteze},
  \citenamefont {DeLuca}, \citenamefont {Minola}, \citenamefont {Marr{\'e}},
  \citenamefont {Gadaleta}, \citenamefont {Brune}, \citenamefont {Nolting},
  \citenamefont {Brookes},\ and\ \citenamefont {Ghiringhelli}}]{sal13}%
  \BibitemOpen
  \bibfield  {author} {\bibinfo {author} {\bibfnamefont {M.}~\bibnamefont
  {Salluzzo}}, \bibinfo {author} {\bibfnamefont {S.}~\bibnamefont {Gariglio}},
  \bibinfo {author} {\bibfnamefont {D.}~\bibnamefont {Stornaiuolo}}, \bibinfo
  {author} {\bibfnamefont {V.}~\bibnamefont {Sessi}}, \bibinfo {author}
  {\bibfnamefont {S.}~\bibnamefont {Rusponi}}, \bibinfo {author} {\bibfnamefont
  {C.}~\bibnamefont {Piamonteze}}, \bibinfo {author} {\bibfnamefont {G.~M.}\
  \bibnamefont {DeLuca}}, \bibinfo {author} {\bibfnamefont {M.}~\bibnamefont
  {Minola}}, \bibinfo {author} {\bibfnamefont {D.}~\bibnamefont {Marr{\'e}}},
  \bibinfo {author} {\bibfnamefont {A.}~\bibnamefont {Gadaleta}}, \bibinfo
  {author} {\bibfnamefont {H.}~\bibnamefont {Brune}}, \bibinfo {author}
  {\bibfnamefont {F.}~\bibnamefont {Nolting}}, \bibinfo {author} {\bibfnamefont
  {N.~B.}\ \bibnamefont {Brookes}}, \ and\ \bibinfo {author} {\bibfnamefont
  {G.}~\bibnamefont {Ghiringhelli}},\ }\href@noop {} {\bibfield  {journal}
  {\bibinfo  {journal} {Phys. Rev. Lett.}\ }\textbf {\bibinfo {volume} {111}},\
  \bibinfo {pages} {087204} (\bibinfo {year} {2013})}\BibitemShut {NoStop}%
\bibitem [{\citenamefont {Liu}\ \emph {et~al.}(2013)\citenamefont {Liu},
  \citenamefont {Li}, \citenamefont {L{\"u}}, \citenamefont {Huang},
  \citenamefont {Huang}, \citenamefont {Zeng}, \citenamefont {Qiu},
  \citenamefont {Huang}, \citenamefont {Annadi}, \citenamefont {Chen},
  \citenamefont {Coey}, \citenamefont {Venkatesan},\ and\ \citenamefont
  {Ariando}}]{liu13}%
  \BibitemOpen
  \bibfield  {author} {\bibinfo {author} {\bibfnamefont {Z.~Q.}\ \bibnamefont
  {Liu}}, \bibinfo {author} {\bibfnamefont {C.~J.}\ \bibnamefont {Li}},
  \bibinfo {author} {\bibfnamefont {W.~M.}\ \bibnamefont {L{\"u}}}, \bibinfo
  {author} {\bibfnamefont {X.~H.}\ \bibnamefont {Huang}}, \bibinfo {author}
  {\bibfnamefont {Z.}~\bibnamefont {Huang}}, \bibinfo {author} {\bibfnamefont
  {S.~W.}\ \bibnamefont {Zeng}}, \bibinfo {author} {\bibfnamefont {X.~P.}\
  \bibnamefont {Qiu}}, \bibinfo {author} {\bibfnamefont {L.~S.}\ \bibnamefont
  {Huang}}, \bibinfo {author} {\bibfnamefont {A.}~\bibnamefont {Annadi}},
  \bibinfo {author} {\bibfnamefont {J.~S.}\ \bibnamefont {Chen}}, \bibinfo
  {author} {\bibfnamefont {J.~M.~D.}\ \bibnamefont {Coey}}, \bibinfo {author}
  {\bibfnamefont {T.}~\bibnamefont {Venkatesan}}, \ and\ \bibinfo {author}
  {\bibnamefont {Ariando}},\ }\href@noop {} {\bibfield  {journal} {\bibinfo
  {journal} {Physical Review X}\ }\textbf {\bibinfo {volume} {3}},\ \bibinfo
  {pages} {021010} (\bibinfo {year} {2013})}\BibitemShut {NoStop}%
\bibitem [{\citenamefont {David}\ \emph {et~al.}(2015)\citenamefont {David},
  \citenamefont {Tian}, \citenamefont {Yang}, \citenamefont {Gao},
  \citenamefont {Lin}, \citenamefont {amd J.-M.~Zuo}, \citenamefont
  {Prellier},\ and\ \citenamefont {Wu}}]{dav15}%
  \BibitemOpen
  \bibfield  {author} {\bibinfo {author} {\bibfnamefont {A.}~\bibnamefont
  {David}}, \bibinfo {author} {\bibfnamefont {Y.}~\bibnamefont {Tian}},
  \bibinfo {author} {\bibfnamefont {P.}~\bibnamefont {Yang}}, \bibinfo {author}
  {\bibfnamefont {X.}~\bibnamefont {Gao}}, \bibinfo {author} {\bibfnamefont
  {W.}~\bibnamefont {Lin}}, \bibinfo {author} {\bibfnamefont {A.~B.~S.}\
  \bibnamefont {amd J.-M.~Zuo}}, \bibinfo {author} {\bibfnamefont
  {W.}~\bibnamefont {Prellier}}, \ and\ \bibinfo {author} {\bibfnamefont
  {T.}~\bibnamefont {Wu}},\ }\href@noop {} {\bibfield  {journal} {\bibinfo
  {journal} {Scientific Reports}\ }\textbf {\bibinfo {volume} {5}},\ \bibinfo
  {pages} {10255} (\bibinfo {year} {2015})}\BibitemShut {NoStop}%
\bibitem [{\citenamefont {Santander-Syro}\ \emph {et~al.}(2011)\citenamefont
  {Santander-Syro}, \citenamefont {Copie}, \citenamefont {Kondo}, \citenamefont
  {Fortuna}, \citenamefont {Pailh{\`e}}, \citenamefont {Weht}, \citenamefont
  {Qiu}, \citenamefont {Bertran}, \citenamefont {Nicolaou}, \citenamefont
  {Taleb-Ibrahimi}, \citenamefont {F{\`e}vre}, \citenamefont {Herranz},
  \citenamefont {Bibes}, \citenamefont {Reyren}, \citenamefont {Apertet},
  \citenamefont {Lecoeur}, \citenamefont {Barth{\'e}l{\'e}my},\ and\
  \citenamefont {Rozenberg}}]{san11}%
  \BibitemOpen
  \bibfield  {author} {\bibinfo {author} {\bibfnamefont {A.~F.}\ \bibnamefont
  {Santander-Syro}}, \bibinfo {author} {\bibfnamefont {O.}~\bibnamefont
  {Copie}}, \bibinfo {author} {\bibfnamefont {T.}~\bibnamefont {Kondo}},
  \bibinfo {author} {\bibfnamefont {F.}~\bibnamefont {Fortuna}}, \bibinfo
  {author} {\bibfnamefont {S.}~\bibnamefont {Pailh{\`e}}}, \bibinfo {author}
  {\bibfnamefont {R.}~\bibnamefont {Weht}}, \bibinfo {author} {\bibfnamefont
  {X.~G.}\ \bibnamefont {Qiu}}, \bibinfo {author} {\bibfnamefont
  {F.}~\bibnamefont {Bertran}}, \bibinfo {author} {\bibfnamefont
  {A.}~\bibnamefont {Nicolaou}}, \bibinfo {author} {\bibfnamefont
  {A.}~\bibnamefont {Taleb-Ibrahimi}}, \bibinfo {author} {\bibfnamefont
  {P.~L.}\ \bibnamefont {F{\`e}vre}}, \bibinfo {author} {\bibfnamefont
  {G.}~\bibnamefont {Herranz}}, \bibinfo {author} {\bibfnamefont
  {M.}~\bibnamefont {Bibes}}, \bibinfo {author} {\bibfnamefont
  {N.}~\bibnamefont {Reyren}}, \bibinfo {author} {\bibfnamefont
  {Y.}~\bibnamefont {Apertet}}, \bibinfo {author} {\bibfnamefont
  {P.}~\bibnamefont {Lecoeur}}, \bibinfo {author} {\bibfnamefont
  {A.}~\bibnamefont {Barth{\'e}l{\'e}my}}, \ and\ \bibinfo {author}
  {\bibfnamefont {M.~J.}\ \bibnamefont {Rozenberg}},\ }\href@noop {} {\bibfield
   {journal} {\bibinfo  {journal} {Nature}\ }\textbf {\bibinfo {volume}
  {469}},\ \bibinfo {pages} {189} (\bibinfo {year} {2011})}\BibitemShut
  {NoStop}%
\bibitem [{\citenamefont {Meevasana}\ \emph {et~al.}(2011)\citenamefont
  {Meevasana}, \citenamefont {King}, \citenamefont {He}, \citenamefont {Mo},
  \citenamefont {Hashimoto}, \citenamefont {Tamai}, \citenamefont
  {Songsiriritthigul}, \citenamefont {Baumberger},\ and\ \citenamefont
  {Shen}}]{mee11}%
  \BibitemOpen
  \bibfield  {author} {\bibinfo {author} {\bibfnamefont {W.}~\bibnamefont
  {Meevasana}}, \bibinfo {author} {\bibfnamefont {P.~D.~C.}\ \bibnamefont
  {King}}, \bibinfo {author} {\bibfnamefont {R.~H.}\ \bibnamefont {He}},
  \bibinfo {author} {\bibfnamefont {S.-K.}\ \bibnamefont {Mo}}, \bibinfo
  {author} {\bibfnamefont {M.}~\bibnamefont {Hashimoto}}, \bibinfo {author}
  {\bibfnamefont {A.}~\bibnamefont {Tamai}}, \bibinfo {author} {\bibfnamefont
  {P.}~\bibnamefont {Songsiriritthigul}}, \bibinfo {author} {\bibfnamefont
  {F.}~\bibnamefont {Baumberger}}, \ and\ \bibinfo {author} {\bibfnamefont
  {Z.-X.}\ \bibnamefont {Shen}},\ }\href@noop {} {\bibfield  {journal}
  {\bibinfo  {journal} {Nat. Mat.}\ }\textbf {\bibinfo {volume} {10}},\
  \bibinfo {pages} {114} (\bibinfo {year} {2011})}\BibitemShut {NoStop}%
\bibitem [{\citenamefont {Walker}\ \emph {et~al.}(2014)\citenamefont {Walker},
  \citenamefont {de~la Torre}, \citenamefont {Bruno}, \citenamefont {Tamai},
  \citenamefont {Kim}, \citenamefont {Hoesch}, \citenamefont {Shi},
  \citenamefont {Bahramy}, \citenamefont {King},\ and\ \citenamefont
  {Baumberger}}]{mck14}%
  \BibitemOpen
  \bibfield  {author} {\bibinfo {author} {\bibfnamefont {S.~M.}\ \bibnamefont
  {Walker}}, \bibinfo {author} {\bibfnamefont {A.}~\bibnamefont {de~la Torre}},
  \bibinfo {author} {\bibfnamefont {F.~Y.}\ \bibnamefont {Bruno}}, \bibinfo
  {author} {\bibfnamefont {A.}~\bibnamefont {Tamai}}, \bibinfo {author}
  {\bibfnamefont {T.~K.}\ \bibnamefont {Kim}}, \bibinfo {author} {\bibfnamefont
  {M.}~\bibnamefont {Hoesch}}, \bibinfo {author} {\bibfnamefont
  {M.}~\bibnamefont {Shi}}, \bibinfo {author} {\bibfnamefont {M.~S.}\
  \bibnamefont {Bahramy}}, \bibinfo {author} {\bibfnamefont {P.~C.}\
  \bibnamefont {King}}, \ and\ \bibinfo {author} {\bibfnamefont
  {F.}~\bibnamefont {Baumberger}},\ }\href@noop {} {\bibfield  {journal}
  {\bibinfo  {journal} {Phys. Rev. Lett.}\ }\textbf {\bibinfo {volume} {113}},\
  \bibinfo {pages} {177601} (\bibinfo {year} {2014})}\BibitemShut {NoStop}%
\bibitem [{\citenamefont {Rice}\ \emph {et~al.}(2014)\citenamefont {Rice},
  \citenamefont {Ambwani}, \citenamefont {Bombeck}, \citenamefont {Thompson},
  \citenamefont {Haugstad}, \citenamefont {Leighton},\ and\ \citenamefont
  {Crooker}}]{ric14}%
  \BibitemOpen
  \bibfield  {author} {\bibinfo {author} {\bibfnamefont {W.~D.}\ \bibnamefont
  {Rice}}, \bibinfo {author} {\bibfnamefont {P.}~\bibnamefont {Ambwani}},
  \bibinfo {author} {\bibfnamefont {M.}~\bibnamefont {Bombeck}}, \bibinfo
  {author} {\bibfnamefont {J.~D.}\ \bibnamefont {Thompson}}, \bibinfo {author}
  {\bibfnamefont {G.}~\bibnamefont {Haugstad}}, \bibinfo {author}
  {\bibfnamefont {C.}~\bibnamefont {Leighton}}, \ and\ \bibinfo {author}
  {\bibfnamefont {S.~A.}\ \bibnamefont {Crooker}},\ }\href@noop {} {\bibfield
  {journal} {\bibinfo  {journal} {Nature Mat.}\ }\textbf {\bibinfo {volume}
  {13}},\ \bibinfo {pages} {481} (\bibinfo {year} {2014})}\BibitemShut
  {NoStop}%
\bibitem [{\citenamefont {Reyren}\ \emph {et~al.}(2007)\citenamefont {Reyren},
  \citenamefont {Thiel}, \citenamefont {Caviglia}, \citenamefont {Kourkoutis},
  \citenamefont {Hammerl}, \citenamefont {Richter}, \citenamefont {Schneider},
  \citenamefont {Kopp}, \citenamefont {R{\"u}etschi}, \citenamefont {Jaccard},
  \citenamefont {Gabay}, \citenamefont {Muller}, \citenamefont {Triscone},\
  and\ \citenamefont {Mannhart}}]{rey07}%
  \BibitemOpen
  \bibfield  {author} {\bibinfo {author} {\bibfnamefont {N.}~\bibnamefont
  {Reyren}}, \bibinfo {author} {\bibfnamefont {S.}~\bibnamefont {Thiel}},
  \bibinfo {author} {\bibfnamefont {A.~D.}\ \bibnamefont {Caviglia}}, \bibinfo
  {author} {\bibfnamefont {L.~F.}\ \bibnamefont {Kourkoutis}}, \bibinfo
  {author} {\bibfnamefont {G.}~\bibnamefont {Hammerl}}, \bibinfo {author}
  {\bibfnamefont {C.}~\bibnamefont {Richter}}, \bibinfo {author} {\bibfnamefont
  {C.~W.}\ \bibnamefont {Schneider}}, \bibinfo {author} {\bibfnamefont
  {T.}~\bibnamefont {Kopp}}, \bibinfo {author} {\bibfnamefont {A.-S.}\
  \bibnamefont {R{\"u}etschi}}, \bibinfo {author} {\bibfnamefont
  {D.}~\bibnamefont {Jaccard}}, \bibinfo {author} {\bibfnamefont
  {M.}~\bibnamefont {Gabay}}, \bibinfo {author} {\bibfnamefont {D.~A.}\
  \bibnamefont {Muller}}, \bibinfo {author} {\bibfnamefont {J.-M.}\
  \bibnamefont {Triscone}}, \ and\ \bibinfo {author} {\bibfnamefont
  {J.}~\bibnamefont {Mannhart}},\ }\href@noop {} {\bibfield  {journal}
  {\bibinfo  {journal} {Science}\ }\textbf {\bibinfo {volume} {317}},\ \bibinfo
  {pages} {1196} (\bibinfo {year} {2007})}\BibitemShut {NoStop}%
\bibitem [{\citenamefont {Brinkman}\ \emph {et~al.}(2007)\citenamefont
  {Brinkman}, \citenamefont {Huijben}, \citenamefont {van Zalk}, \citenamefont
  {Huijben}, \citenamefont {Zeitler}, \citenamefont {Maan}, \citenamefont
  {van~der Wiel}, \citenamefont {Rijnders}, \citenamefont {Blank},\ and\
  \citenamefont {Hilgenkamp}}]{bri07}%
  \BibitemOpen
  \bibfield  {author} {\bibinfo {author} {\bibfnamefont {A.}~\bibnamefont
  {Brinkman}}, \bibinfo {author} {\bibfnamefont {M.}~\bibnamefont {Huijben}},
  \bibinfo {author} {\bibfnamefont {M.}~\bibnamefont {van Zalk}}, \bibinfo
  {author} {\bibfnamefont {J.}~\bibnamefont {Huijben}}, \bibinfo {author}
  {\bibfnamefont {U.}~\bibnamefont {Zeitler}}, \bibinfo {author} {\bibfnamefont
  {J.~C.}\ \bibnamefont {Maan}}, \bibinfo {author} {\bibfnamefont {W.~G.}\
  \bibnamefont {van~der Wiel}}, \bibinfo {author} {\bibfnamefont
  {G.}~\bibnamefont {Rijnders}}, \bibinfo {author} {\bibfnamefont {D.~H.~A.}\
  \bibnamefont {Blank}}, \ and\ \bibinfo {author} {\bibfnamefont
  {H.}~\bibnamefont {Hilgenkamp}},\ }\href@noop {} {\bibfield  {journal}
  {\bibinfo  {journal} {Nature Mater.}\ }\textbf {\bibinfo {volume} {6}},\
  \bibinfo {pages} {493} (\bibinfo {year} {2007})}\BibitemShut {NoStop}%
\bibitem [{\citenamefont {Li}\ \emph {et~al.}(2011)\citenamefont {Li},
  \citenamefont {Phattalung}, \citenamefont {Limpijumnong}, \citenamefont
  {Kim},\ and\ \citenamefont {Yu}}]{li11}%
  \BibitemOpen
  \bibfield  {author} {\bibinfo {author} {\bibfnamefont {Y.}~\bibnamefont
  {Li}}, \bibinfo {author} {\bibfnamefont {S.~N.}\ \bibnamefont {Phattalung}},
  \bibinfo {author} {\bibfnamefont {S.}~\bibnamefont {Limpijumnong}}, \bibinfo
  {author} {\bibfnamefont {J.}~\bibnamefont {Kim}}, \ and\ \bibinfo {author}
  {\bibfnamefont {J.}~\bibnamefont {Yu}},\ }\href@noop {} {\bibfield  {journal}
  {\bibinfo  {journal} {Phys. Rev. B}\ }\textbf {\bibinfo {volume} {84}},\
  \bibinfo {pages} {245307} (\bibinfo {year} {2011})}\BibitemShut {NoStop}%
\bibitem [{\citenamefont {Ariando}\ \emph {et~al.}(2011)\citenamefont
  {Ariando}, \citenamefont {Wang}, \citenamefont {Baskaran}, \citenamefont
  {Liu}, \citenamefont {Huijben}, \citenamefont {Yi}, \citenamefont {Annadi},
  \citenamefont {Barman}, \citenamefont {Rusydi}, \citenamefont {Dhar},
  \citenamefont {Feng}, \citenamefont {Ding}, \citenamefont {Hilgenkamp},\ and\
  \citenamefont {Venkatesan}}]{ari11}%
  \BibitemOpen
  \bibfield  {author} {\bibinfo {author} {\bibnamefont {Ariando}}, \bibinfo
  {author} {\bibfnamefont {X.}~\bibnamefont {Wang}}, \bibinfo {author}
  {\bibfnamefont {G.}~\bibnamefont {Baskaran}}, \bibinfo {author}
  {\bibfnamefont {Z.~Q.}\ \bibnamefont {Liu}}, \bibinfo {author} {\bibfnamefont
  {J.}~\bibnamefont {Huijben}}, \bibinfo {author} {\bibfnamefont {J.~B.}\
  \bibnamefont {Yi}}, \bibinfo {author} {\bibfnamefont {A.}~\bibnamefont
  {Annadi}}, \bibinfo {author} {\bibfnamefont {A.~R.}\ \bibnamefont {Barman}},
  \bibinfo {author} {\bibfnamefont {A.}~\bibnamefont {Rusydi}}, \bibinfo
  {author} {\bibfnamefont {S.}~\bibnamefont {Dhar}}, \bibinfo {author}
  {\bibfnamefont {Y.~P.}\ \bibnamefont {Feng}}, \bibinfo {author}
  {\bibfnamefont {J.}~\bibnamefont {Ding}}, \bibinfo {author} {\bibfnamefont
  {H.}~\bibnamefont {Hilgenkamp}}, \ and\ \bibinfo {author} {\bibfnamefont
  {T.}~\bibnamefont {Venkatesan}},\ }\href@noop {} {\bibfield  {journal}
  {\bibinfo  {journal} {Nature Commun.}\ }\textbf {\bibinfo {volume} {2}},\
  \bibinfo {pages} {188} (\bibinfo {year} {2011})}\BibitemShut {NoStop}%
\bibitem [{\citenamefont {Lee}\ \emph {et~al.}(2013)\citenamefont {Lee},
  \citenamefont {Xie}, \citenamefont {Sato}, \citenamefont {Bell},
  \citenamefont {Hikita}, \citenamefont {Hwang},\ and\ \citenamefont
  {Kao}}]{lee13}%
  \BibitemOpen
  \bibfield  {author} {\bibinfo {author} {\bibfnamefont {J.-S.}\ \bibnamefont
  {Lee}}, \bibinfo {author} {\bibfnamefont {Y.~W.}\ \bibnamefont {Xie}},
  \bibinfo {author} {\bibfnamefont {H.~K.}\ \bibnamefont {Sato}}, \bibinfo
  {author} {\bibfnamefont {C.}~\bibnamefont {Bell}}, \bibinfo {author}
  {\bibfnamefont {Y.}~\bibnamefont {Hikita}}, \bibinfo {author} {\bibfnamefont
  {H.~Y.}\ \bibnamefont {Hwang}}, \ and\ \bibinfo {author} {\bibfnamefont
  {C.-C.}\ \bibnamefont {Kao}},\ }\href@noop {} {\bibfield  {journal} {\bibinfo
   {journal} {Nature Mat.}\ }\textbf {\bibinfo {volume} {12}},\ \bibinfo
  {pages} {703} (\bibinfo {year} {2013})}\BibitemShut {NoStop}%
\bibitem [{\citenamefont {Pentcheva}\ and\ \citenamefont
  {Pickett}(2006)}]{pen06}%
  \BibitemOpen
  \bibfield  {author} {\bibinfo {author} {\bibfnamefont {R.}~\bibnamefont
  {Pentcheva}}\ and\ \bibinfo {author} {\bibfnamefont {W.~E.}\ \bibnamefont
  {Pickett}},\ }\href@noop {} {\bibfield  {journal} {\bibinfo  {journal} {Phys.
  Rev. B}\ }\textbf {\bibinfo {volume} {74}},\ \bibinfo {pages} {035112}
  (\bibinfo {year} {2006})}\BibitemShut {NoStop}%
\bibitem [{\citenamefont {Pavlenko}\ \emph {et~al.}(2012)\citenamefont
  {Pavlenko}, \citenamefont {Kopp}, \citenamefont {Tsymbal}, \citenamefont
  {Sawatzky},\ and\ \citenamefont {Mannhart}}]{pav12}%
  \BibitemOpen
  \bibfield  {author} {\bibinfo {author} {\bibfnamefont {N.}~\bibnamefont
  {Pavlenko}}, \bibinfo {author} {\bibfnamefont {T.}~\bibnamefont {Kopp}},
  \bibinfo {author} {\bibfnamefont {E.~Y.}\ \bibnamefont {Tsymbal}}, \bibinfo
  {author} {\bibfnamefont {G.~A.}\ \bibnamefont {Sawatzky}}, \ and\ \bibinfo
  {author} {\bibfnamefont {J.}~\bibnamefont {Mannhart}},\ }\href@noop {}
  {\bibfield  {journal} {\bibinfo  {journal} {Phys. Rev. B}\ }\textbf {\bibinfo
  {volume} {85}},\ \bibinfo {pages} {020407(R)} (\bibinfo {year}
  {2012})}\BibitemShut {NoStop}%
\bibitem [{\citenamefont {Shen}\ \emph {et~al.}(2012)\citenamefont {Shen},
  \citenamefont {Lee}, \citenamefont {Valent{\'i}},\ and\ \citenamefont
  {Jeschke}}]{she12}%
  \BibitemOpen
  \bibfield  {author} {\bibinfo {author} {\bibfnamefont {J.}~\bibnamefont
  {Shen}}, \bibinfo {author} {\bibfnamefont {H.}~\bibnamefont {Lee}}, \bibinfo
  {author} {\bibfnamefont {R.}~\bibnamefont {Valent{\'i}}}, \ and\ \bibinfo
  {author} {\bibfnamefont {H.~O.}\ \bibnamefont {Jeschke}},\ }\href@noop {}
  {\bibfield  {journal} {\bibinfo  {journal} {Phys. Rev. B}\ }\textbf {\bibinfo
  {volume} {86}},\ \bibinfo {pages} {195119} (\bibinfo {year}
  {2012})}\BibitemShut {NoStop}%
\bibitem [{\citenamefont {Lin}\ and\ \citenamefont {Demkov}(2013)}]{lin13}%
  \BibitemOpen
  \bibfield  {author} {\bibinfo {author} {\bibfnamefont {C.}~\bibnamefont
  {Lin}}\ and\ \bibinfo {author} {\bibfnamefont {A.~A.}\ \bibnamefont
  {Demkov}},\ }\href@noop {} {\bibfield  {journal} {\bibinfo  {journal} {Phys.
  Rev. Lett.}\ }\textbf {\bibinfo {volume} {111}},\ \bibinfo {pages} {217601}
  (\bibinfo {year} {2013})}\BibitemShut {NoStop}%
\bibitem [{\citenamefont {Lechermann}\ \emph {et~al.}(2014)\citenamefont
  {Lechermann}, \citenamefont {Boehnke}, \citenamefont {Grieger},\ and\
  \citenamefont {Piefke}}]{lec14}%
  \BibitemOpen
  \bibfield  {author} {\bibinfo {author} {\bibfnamefont {F.}~\bibnamefont
  {Lechermann}}, \bibinfo {author} {\bibfnamefont {L.}~\bibnamefont {Boehnke}},
  \bibinfo {author} {\bibfnamefont {D.}~\bibnamefont {Grieger}}, \ and\
  \bibinfo {author} {\bibfnamefont {C.}~\bibnamefont {Piefke}},\ }\href@noop {}
  {\bibfield  {journal} {\bibinfo  {journal} {Phys. Rev. B}\ }\textbf {\bibinfo
  {volume} {90}},\ \bibinfo {pages} {085125} (\bibinfo {year}
  {2014})}\BibitemShut {NoStop}%
\bibitem [{\citenamefont {Breitschaft}\ \emph {et~al.}(2010)\citenamefont
  {Breitschaft}, \citenamefont {Tinkl}, \citenamefont {Pavlenko}, \citenamefont
  {Paetel}, \citenamefont {Richter}, \citenamefont {Kirtley}, \citenamefont
  {Liao}, \citenamefont {Hammerl}, \citenamefont {Eyert}, \citenamefont
  {Kopp},\ and\ \citenamefont {Mannhart}}]{bre10}%
  \BibitemOpen
  \bibfield  {author} {\bibinfo {author} {\bibfnamefont {M.}~\bibnamefont
  {Breitschaft}}, \bibinfo {author} {\bibfnamefont {V.}~\bibnamefont {Tinkl}},
  \bibinfo {author} {\bibfnamefont {N.}~\bibnamefont {Pavlenko}}, \bibinfo
  {author} {\bibfnamefont {S.}~\bibnamefont {Paetel}}, \bibinfo {author}
  {\bibfnamefont {C.}~\bibnamefont {Richter}}, \bibinfo {author} {\bibfnamefont
  {J.~R.}\ \bibnamefont {Kirtley}}, \bibinfo {author} {\bibfnamefont {Y.~C.}\
  \bibnamefont {Liao}}, \bibinfo {author} {\bibfnamefont {G.}~\bibnamefont
  {Hammerl}}, \bibinfo {author} {\bibfnamefont {V.}~\bibnamefont {Eyert}},
  \bibinfo {author} {\bibfnamefont {T.}~\bibnamefont {Kopp}}, \ and\ \bibinfo
  {author} {\bibfnamefont {J.}~\bibnamefont {Mannhart}},\ }\href@noop {}
  {\bibfield  {journal} {\bibinfo  {journal} {Phys. Rev. B}\ }\textbf {\bibinfo
  {volume} {81}},\ \bibinfo {pages} {153414} (\bibinfo {year}
  {2010})}\BibitemShut {NoStop}%
\bibitem [{\citenamefont {Ristic}\ \emph {et~al.}(2012)\citenamefont {Ristic},
  \citenamefont {DiCapua}, \citenamefont {Chiarella}, \citenamefont {DeLuca},
  \citenamefont {Maggio-Aprile}, \citenamefont {Radovic},\ and\ \citenamefont
  {Salluzzo}}]{ris12}%
  \BibitemOpen
  \bibfield  {author} {\bibinfo {author} {\bibfnamefont {Z.}~\bibnamefont
  {Ristic}}, \bibinfo {author} {\bibfnamefont {R.}~\bibnamefont {DiCapua}},
  \bibinfo {author} {\bibfnamefont {F.}~\bibnamefont {Chiarella}}, \bibinfo
  {author} {\bibfnamefont {G.~M.}\ \bibnamefont {DeLuca}}, \bibinfo {author}
  {\bibfnamefont {I.}~\bibnamefont {Maggio-Aprile}}, \bibinfo {author}
  {\bibfnamefont {M.}~\bibnamefont {Radovic}}, \ and\ \bibinfo {author}
  {\bibfnamefont {M.}~\bibnamefont {Salluzzo}},\ }\href@noop {} {\bibfield
  {journal} {\bibinfo  {journal} {Phys. Rev. B}\ }\textbf {\bibinfo {volume}
  {86}},\ \bibinfo {pages} {045127} (\bibinfo {year} {2012})}\BibitemShut
  {NoStop}%
\bibitem [{\citenamefont {Sitaputra}\ \emph {et~al.}(2015)\citenamefont
  {Sitaputra}, \citenamefont {Sivadas}, \citenamefont {Skowronski},
  \citenamefont {Xiao},\ and\ \citenamefont {Feenstra}}]{sit15}%
  \BibitemOpen
  \bibfield  {author} {\bibinfo {author} {\bibfnamefont {W.}~\bibnamefont
  {Sitaputra}}, \bibinfo {author} {\bibfnamefont {N.}~\bibnamefont {Sivadas}},
  \bibinfo {author} {\bibfnamefont {M.}~\bibnamefont {Skowronski}}, \bibinfo
  {author} {\bibfnamefont {D.}~\bibnamefont {Xiao}}, \ and\ \bibinfo {author}
  {\bibfnamefont {R.~M.}\ \bibnamefont {Feenstra}},\ }\href@noop {} {\bibfield
  {journal} {\bibinfo  {journal} {Phys. Rev. B}\ }\textbf {\bibinfo {volume}
  {91}},\ \bibinfo {pages} {205408} (\bibinfo {year} {2015})}\BibitemShut
  {NoStop}%
\bibitem [{\citenamefont {Joshua}\ \emph {et~al.}(2013)\citenamefont {Joshua},
  \citenamefont {Ruhman}, \citenamefont {Pecker}, \citenamefont {Altman},\ and\
  \citenamefont {Ilani}}]{jos13}%
  \BibitemOpen
  \bibfield  {author} {\bibinfo {author} {\bibfnamefont {A.}~\bibnamefont
  {Joshua}}, \bibinfo {author} {\bibfnamefont {J.}~\bibnamefont {Ruhman}},
  \bibinfo {author} {\bibfnamefont {S.}~\bibnamefont {Pecker}}, \bibinfo
  {author} {\bibfnamefont {E.}~\bibnamefont {Altman}}, \ and\ \bibinfo {author}
  {\bibfnamefont {S.}~\bibnamefont {Ilani}},\ }\href@noop {} {\bibfield
  {journal} {\bibinfo  {journal} {PNAS}\ }\textbf {\bibinfo {volume} {110}},\
  \bibinfo {pages} {9633} (\bibinfo {year} {2013})}\BibitemShut {NoStop}%
\bibitem [{\citenamefont {Zhou}\ \emph {et~al.}(2011)\citenamefont {Zhou},
  \citenamefont {Radovic}, \citenamefont {Schlappa}, \citenamefont {Strocov},
  \citenamefont {Frison}, \citenamefont {Mesot}, \citenamefont {Patthey},\ and\
  \citenamefont {Schmitt}}]{zho11}%
  \BibitemOpen
  \bibfield  {author} {\bibinfo {author} {\bibfnamefont {K.-J.}\ \bibnamefont
  {Zhou}}, \bibinfo {author} {\bibfnamefont {M.}~\bibnamefont {Radovic}},
  \bibinfo {author} {\bibfnamefont {J.}~\bibnamefont {Schlappa}}, \bibinfo
  {author} {\bibfnamefont {V.}~\bibnamefont {Strocov}}, \bibinfo {author}
  {\bibfnamefont {R.}~\bibnamefont {Frison}}, \bibinfo {author} {\bibfnamefont
  {J.}~\bibnamefont {Mesot}}, \bibinfo {author} {\bibfnamefont
  {L.}~\bibnamefont {Patthey}}, \ and\ \bibinfo {author} {\bibfnamefont
  {T.}~\bibnamefont {Schmitt}},\ }\href@noop {} {\bibfield  {journal} {\bibinfo
   {journal} {Phys. Rev. B}\ }\textbf {\bibinfo {volume} {83}},\ \bibinfo
  {pages} {201402(R)} (\bibinfo {year} {2011})}\BibitemShut {NoStop}%
\bibitem [{\citenamefont {Park}\ \emph {et~al.}(2013)\citenamefont {Park},
  \citenamefont {Cho}, \citenamefont {Kim}, \citenamefont {Koo}, \citenamefont
  {Jang}, \citenamefont {Ko}, \citenamefont {Park}, \citenamefont {Lee},
  \citenamefont {Kim}, \citenamefont {Lee}, \citenamefont {Burns},
  \citenamefont {Seo},\ and\ \citenamefont {Lee}}]{par13}%
  \BibitemOpen
  \bibfield  {author} {\bibinfo {author} {\bibfnamefont {J.}~\bibnamefont
  {Park}}, \bibinfo {author} {\bibfnamefont {B.-G.}\ \bibnamefont {Cho}},
  \bibinfo {author} {\bibfnamefont {K.~D.}\ \bibnamefont {Kim}}, \bibinfo
  {author} {\bibfnamefont {J.}~\bibnamefont {Koo}}, \bibinfo {author}
  {\bibfnamefont {H.}~\bibnamefont {Jang}}, \bibinfo {author} {\bibfnamefont
  {K.-T.}\ \bibnamefont {Ko}}, \bibinfo {author} {\bibfnamefont {J.-H.}\
  \bibnamefont {Park}}, \bibinfo {author} {\bibfnamefont {K.-B.}\ \bibnamefont
  {Lee}}, \bibinfo {author} {\bibfnamefont {J.-Y.}\ \bibnamefont {Kim}},
  \bibinfo {author} {\bibfnamefont {D.~R.}\ \bibnamefont {Lee}}, \bibinfo
  {author} {\bibfnamefont {C.~A.}\ \bibnamefont {Burns}}, \bibinfo {author}
  {\bibfnamefont {S.~S.~A.}\ \bibnamefont {Seo}}, \ and\ \bibinfo {author}
  {\bibfnamefont {H.~N.}\ \bibnamefont {Lee}},\ }\href@noop {} {\bibfield
  {journal} {\bibinfo  {journal} {Phys. Rev. Lett.}\ }\textbf {\bibinfo
  {volume} {110}},\ \bibinfo {pages} {017401} (\bibinfo {year}
  {2013})}\BibitemShut {NoStop}%
\bibitem [{\citenamefont {Berner}\ \emph {et~al.}(2013)\citenamefont {Berner},
  \citenamefont {Sing}, \citenamefont {Fujiwara}, \citenamefont {Yasui},
  \citenamefont {Saitoh}, \citenamefont {Yamasaki}, \citenamefont {Nishitani},
  \citenamefont {Sekiyama}, \citenamefont {Pavlenko}, \citenamefont {Kopp},
  \citenamefont {Richter}, \citenamefont {Mannhart}, \citenamefont {Suga}, ,\
  and\ \citenamefont {Claessen}}]{ber13}%
  \BibitemOpen
  \bibfield  {author} {\bibinfo {author} {\bibfnamefont {G.}~\bibnamefont
  {Berner}}, \bibinfo {author} {\bibfnamefont {M.}~\bibnamefont {Sing}},
  \bibinfo {author} {\bibfnamefont {H.}~\bibnamefont {Fujiwara}}, \bibinfo
  {author} {\bibfnamefont {A.}~\bibnamefont {Yasui}}, \bibinfo {author}
  {\bibfnamefont {Y.}~\bibnamefont {Saitoh}}, \bibinfo {author} {\bibfnamefont
  {A.}~\bibnamefont {Yamasaki}}, \bibinfo {author} {\bibfnamefont
  {Y.}~\bibnamefont {Nishitani}}, \bibinfo {author} {\bibfnamefont
  {A.}~\bibnamefont {Sekiyama}}, \bibinfo {author} {\bibfnamefont
  {N.}~\bibnamefont {Pavlenko}}, \bibinfo {author} {\bibfnamefont
  {T.}~\bibnamefont {Kopp}}, \bibinfo {author} {\bibfnamefont {C.}~\bibnamefont
  {Richter}}, \bibinfo {author} {\bibfnamefont {J.}~\bibnamefont {Mannhart}},
  \bibinfo {author} {\bibfnamefont {S.}~\bibnamefont {Suga}}, , \ and\ \bibinfo
  {author} {\bibfnamefont {R.}~\bibnamefont {Claessen}},\ }\href@noop {}
  {\bibfield  {journal} {\bibinfo  {journal} {Phys. Rev. Lett.}\ }\textbf
  {\bibinfo {volume} {110}},\ \bibinfo {pages} {247601} (\bibinfo {year}
  {2013})}\BibitemShut {NoStop}%
\bibitem [{\citenamefont {Sulpizio}\ \emph {et~al.}(2014)\citenamefont
  {Sulpizio}, \citenamefont {Ilani}, \citenamefont {Irvin},\ and\ \citenamefont
  {Levy}}]{sul14}%
  \BibitemOpen
  \bibfield  {author} {\bibinfo {author} {\bibfnamefont {J.~A.}\ \bibnamefont
  {Sulpizio}}, \bibinfo {author} {\bibfnamefont {S.}~\bibnamefont {Ilani}},
  \bibinfo {author} {\bibfnamefont {P.}~\bibnamefont {Irvin}}, \ and\ \bibinfo
  {author} {\bibfnamefont {J.}~\bibnamefont {Levy}},\ }\href@noop {} {\bibfield
   {journal} {\bibinfo  {journal} {Annu. Rev. Mater. Res.}\ }\textbf {\bibinfo
  {volume} {44}},\ \bibinfo {pages} {117} (\bibinfo {year} {2014})}\BibitemShut
  {NoStop}%
\bibitem [{\citenamefont {Bi}\ \emph {et~al.}(2014)\citenamefont {Bi},
  \citenamefont {Huang}, \citenamefont {Ryu}, \citenamefont {Lee},
  \citenamefont {Bark}, \citenamefont {amd P.~Irvin},\ and\ \citenamefont
  {Levy}}]{bi14}%
  \BibitemOpen
  \bibfield  {author} {\bibinfo {author} {\bibfnamefont {F.}~\bibnamefont
  {Bi}}, \bibinfo {author} {\bibfnamefont {M.}~\bibnamefont {Huang}}, \bibinfo
  {author} {\bibfnamefont {S.}~\bibnamefont {Ryu}}, \bibinfo {author}
  {\bibfnamefont {H.}~\bibnamefont {Lee}}, \bibinfo {author} {\bibfnamefont
  {C.-W.}\ \bibnamefont {Bark}}, \bibinfo {author} {\bibfnamefont {C.-B.~E.}\
  \bibnamefont {amd P.~Irvin}}, \ and\ \bibinfo {author} {\bibfnamefont
  {J.}~\bibnamefont {Levy}},\ }\href@noop {} {\bibfield  {journal} {\bibinfo
  {journal} {Nature Commun.}\ }\textbf {\bibinfo {volume} {5}},\ \bibinfo
  {pages} {5019} (\bibinfo {year} {2014})}\BibitemShut {NoStop}%
\bibitem [{\citenamefont {Bark}\ \emph {et~al.}(2012)\citenamefont {Bark},
  \citenamefont {Sharma}, \citenamefont {Wang}, \citenamefont {Baek},
  \citenamefont {Lee}, \citenamefont {Ryu}, \citenamefont {Folkman},
  \citenamefont {Paudel}, \citenamefont {Kumar}, \citenamefont {Kalinin},
  \citenamefont {Sokolov}, \citenamefont {Tsymbal}, \citenamefont {Rzchowski},
  \citenamefont {Gruverman},\ and\ \citenamefont {Eom}}]{bar12}%
  \BibitemOpen
  \bibfield  {author} {\bibinfo {author} {\bibfnamefont {C.~W.}\ \bibnamefont
  {Bark}}, \bibinfo {author} {\bibfnamefont {P.}~\bibnamefont {Sharma}},
  \bibinfo {author} {\bibfnamefont {Y.}~\bibnamefont {Wang}}, \bibinfo {author}
  {\bibfnamefont {S.~H.}\ \bibnamefont {Baek}}, \bibinfo {author}
  {\bibfnamefont {S.}~\bibnamefont {Lee}}, \bibinfo {author} {\bibfnamefont
  {S.}~\bibnamefont {Ryu}}, \bibinfo {author} {\bibfnamefont {C.~M.}\
  \bibnamefont {Folkman}}, \bibinfo {author} {\bibfnamefont {T.~R.}\
  \bibnamefont {Paudel}}, \bibinfo {author} {\bibfnamefont {A.}~\bibnamefont
  {Kumar}}, \bibinfo {author} {\bibfnamefont {S.~V.}\ \bibnamefont {Kalinin}},
  \bibinfo {author} {\bibfnamefont {A.}~\bibnamefont {Sokolov}}, \bibinfo
  {author} {\bibfnamefont {E.~Y.}\ \bibnamefont {Tsymbal}}, \bibinfo {author}
  {\bibfnamefont {M.~S.}\ \bibnamefont {Rzchowski}}, \bibinfo {author}
  {\bibfnamefont {A.}~\bibnamefont {Gruverman}}, \ and\ \bibinfo {author}
  {\bibfnamefont {C.~B.}\ \bibnamefont {Eom}},\ }\href@noop {} {\bibfield
  {journal} {\bibinfo  {journal} {Nano Lett.}\ }\textbf {\bibinfo {volume}
  {12}},\ \bibinfo {pages} {1765} (\bibinfo {year} {2012})}\BibitemShut
  {NoStop}%
\bibitem [{\citenamefont {Shein}\ and\ \citenamefont
  {Ivanovskii}(2007)}]{she07}%
  \BibitemOpen
  \bibfield  {author} {\bibinfo {author} {\bibfnamefont {I.~R.}\ \bibnamefont
  {Shein}}\ and\ \bibinfo {author} {\bibfnamefont {A.~L.}\ \bibnamefont
  {Ivanovskii}},\ }\href@noop {} {\bibfield  {journal} {\bibinfo  {journal}
  {Phys. Lett. A}\ }\textbf {\bibinfo {volume} {371}},\ \bibinfo {pages} {155}
  (\bibinfo {year} {2007})}\BibitemShut {NoStop}%
\bibitem [{\citenamefont {Johnson-Wilke}\ \emph {et~al.}(2013)\citenamefont
  {Johnson-Wilke}, \citenamefont {Marincel}, \citenamefont {Zhu}, \citenamefont
  {Warusawithana}, \citenamefont {Hatt}, \citenamefont {Sayre}, \citenamefont
  {Delaney}, \citenamefont {Engel-Herbert}, \citenamefont {Schlep\"utz},
  \citenamefont {Kim}, \citenamefont {Gopalan}, \citenamefont {Spaldin},
  \citenamefont {Schlom}, \citenamefont {Ryan},\ and\ \citenamefont
  {Trolier-McKinstry}}]{joh12}%
  \BibitemOpen
  \bibfield  {author} {\bibinfo {author} {\bibfnamefont {R.~L.}\ \bibnamefont
  {Johnson-Wilke}}, \bibinfo {author} {\bibfnamefont {D.}~\bibnamefont
  {Marincel}}, \bibinfo {author} {\bibfnamefont {S.}~\bibnamefont {Zhu}},
  \bibinfo {author} {\bibfnamefont {M.~P.}\ \bibnamefont {Warusawithana}},
  \bibinfo {author} {\bibfnamefont {A.}~\bibnamefont {Hatt}}, \bibinfo {author}
  {\bibfnamefont {J.}~\bibnamefont {Sayre}}, \bibinfo {author} {\bibfnamefont
  {K.~T.}\ \bibnamefont {Delaney}}, \bibinfo {author} {\bibfnamefont
  {R.}~\bibnamefont {Engel-Herbert}}, \bibinfo {author} {\bibfnamefont {C.~M.}\
  \bibnamefont {Schlep\"utz}}, \bibinfo {author} {\bibfnamefont {J.-W.}\
  \bibnamefont {Kim}}, \bibinfo {author} {\bibfnamefont {V.}~\bibnamefont
  {Gopalan}}, \bibinfo {author} {\bibfnamefont {N.~A.}\ \bibnamefont
  {Spaldin}}, \bibinfo {author} {\bibfnamefont {D.~G.}\ \bibnamefont {Schlom}},
  \bibinfo {author} {\bibfnamefont {P.~J.}\ \bibnamefont {Ryan}}, \ and\
  \bibinfo {author} {\bibfnamefont {S.}~\bibnamefont {Trolier-McKinstry}},\
  }\href@noop {} {\bibfield  {journal} {\bibinfo  {journal} {Phys. Rev. B}\
  }\textbf {\bibinfo {volume} {88}},\ \bibinfo {pages} {174101} (\bibinfo
  {year} {2013})}\BibitemShut {NoStop}%
\bibitem [{\citenamefont {Li}\ \emph {et~al.}(2013)\citenamefont {Li},
  \citenamefont {Beltr{\'a}n},\ and\ \citenamefont {Mu{\~n}oz}}]{li13}%
  \BibitemOpen
  \bibfield  {author} {\bibinfo {author} {\bibfnamefont {J.~C.}\ \bibnamefont
  {Li}}, \bibinfo {author} {\bibfnamefont {J.~I.}\ \bibnamefont {Beltr{\'a}n}},
  \ and\ \bibinfo {author} {\bibfnamefont {M.~C.}\ \bibnamefont {Mu{\~n}oz}},\
  }\href@noop {} {\bibfield  {journal} {\bibinfo  {journal} {Phys. Rev. B}\
  }\textbf {\bibinfo {volume} {87}},\ \bibinfo {pages} {075411} (\bibinfo
  {year} {2013})}\BibitemShut {NoStop}%
\bibitem [{\citenamefont {Janotti}\ \emph {et~al.}(2014)\citenamefont
  {Janotti}, \citenamefont {Varley}, \citenamefont {Choi},\ and\ \citenamefont
  {Van~de Walle}}]{jan14}%
  \BibitemOpen
  \bibfield  {author} {\bibinfo {author} {\bibfnamefont {A.}~\bibnamefont
  {Janotti}}, \bibinfo {author} {\bibfnamefont {J.~B.}\ \bibnamefont {Varley}},
  \bibinfo {author} {\bibfnamefont {M.}~\bibnamefont {Choi}}, \ and\ \bibinfo
  {author} {\bibfnamefont {C.~G.}\ \bibnamefont {Van~de Walle}},\ }\href@noop
  {} {\bibfield  {journal} {\bibinfo  {journal} {Phys. Rev. B}\ }\textbf
  {\bibinfo {volume} {90}},\ \bibinfo {pages} {085202} (\bibinfo {year}
  {2014})}\BibitemShut {NoStop}%
\bibitem [{\citenamefont {Lopez-Bezanilla}\ \emph {et~al.}(2015)\citenamefont
  {Lopez-Bezanilla}, \citenamefont {Ganesh},\ and\ \citenamefont
  {Littlewood}}]{lop15}%
  \BibitemOpen
  \bibfield  {author} {\bibinfo {author} {\bibfnamefont {A.}~\bibnamefont
  {Lopez-Bezanilla}}, \bibinfo {author} {\bibfnamefont {P.}~\bibnamefont
  {Ganesh}}, \ and\ \bibinfo {author} {\bibfnamefont {P.~B.}\ \bibnamefont
  {Littlewood}},\ }\href@noop {} {\bibfield  {journal} {\bibinfo  {journal}
  {Phys. Rev. B}\ }\textbf {\bibinfo {volume} {92}},\ \bibinfo {pages} {115112}
  (\bibinfo {year} {2015})}\BibitemShut {NoStop}%
\bibitem [{\citenamefont {Grieger}\ and\ \citenamefont
  {Lechermann}(2014)}]{gri14}%
  \BibitemOpen
  \bibfield  {author} {\bibinfo {author} {\bibfnamefont {D.}~\bibnamefont
  {Grieger}}\ and\ \bibinfo {author} {\bibfnamefont {F.}~\bibnamefont
  {Lechermann}},\ }\href@noop {} {\bibfield  {journal} {\bibinfo  {journal}
  {Phys. Rev. B}\ }\textbf {\bibinfo {volume} {90}},\ \bibinfo {pages} {115115}
  (\bibinfo {year} {2014})}\BibitemShut {NoStop}%
\bibitem [{\citenamefont {Michaeli}\ \emph {et~al.}(2012)\citenamefont
  {Michaeli}, \citenamefont {Potter},\ and\ \citenamefont {Lee}}]{mic12}%
  \BibitemOpen
  \bibfield  {author} {\bibinfo {author} {\bibfnamefont {K.}~\bibnamefont
  {Michaeli}}, \bibinfo {author} {\bibfnamefont {A.~C.}\ \bibnamefont
  {Potter}}, \ and\ \bibinfo {author} {\bibfnamefont {P.~A.}\ \bibnamefont
  {Lee}},\ }\href@noop {} {\bibfield  {journal} {\bibinfo  {journal} {Phys.
  Rev. Lett.}\ }\textbf {\bibinfo {volume} {108}},\ \bibinfo {pages} {117003}
  (\bibinfo {year} {2012})}\BibitemShut {NoStop}%
\bibitem [{\citenamefont {Chen}\ and\ \citenamefont {Balents}(2013)}]{che13_2}%
  \BibitemOpen
  \bibfield  {author} {\bibinfo {author} {\bibfnamefont {G.}~\bibnamefont
  {Chen}}\ and\ \bibinfo {author} {\bibfnamefont {L.}~\bibnamefont {Balents}},\
  }\href@noop {} {\bibfield  {journal} {\bibinfo  {journal} {Phys. Rev. Lett.}\
  }\textbf {\bibinfo {volume} {110}},\ \bibinfo {pages} {206401} (\bibinfo
  {year} {2013})}\BibitemShut {NoStop}%
\bibitem [{\citenamefont {Banerjee}\ \emph {et~al.}(2013)\citenamefont
  {Banerjee}, \citenamefont {Erten},\ and\ \citenamefont {Randeria}}]{ban13}%
  \BibitemOpen
  \bibfield  {author} {\bibinfo {author} {\bibfnamefont {S.}~\bibnamefont
  {Banerjee}}, \bibinfo {author} {\bibfnamefont {O.}~\bibnamefont {Erten}}, \
  and\ \bibinfo {author} {\bibfnamefont {M.}~\bibnamefont {Randeria}},\
  }\href@noop {} {\bibfield  {journal} {\bibinfo  {journal} {Nature Phys.}\
  }\textbf {\bibinfo {volume} {9}},\ \bibinfo {pages} {626} (\bibinfo {year}
  {2013})}\BibitemShut {NoStop}%
\bibitem [{\citenamefont {Ruhman}\ \emph {et~al.}(2013)\citenamefont {Ruhman},
  \citenamefont {Joshua}, \citenamefont {Ilani},\ and\ \citenamefont
  {Altman}}]{ruh13}%
  \BibitemOpen
  \bibfield  {author} {\bibinfo {author} {\bibfnamefont {J.}~\bibnamefont
  {Ruhman}}, \bibinfo {author} {\bibfnamefont {A.}~\bibnamefont {Joshua}},
  \bibinfo {author} {\bibfnamefont {S.}~\bibnamefont {Ilani}}, \ and\ \bibinfo
  {author} {\bibfnamefont {E.}~\bibnamefont {Altman}},\ }\href@noop {}
  {\bibfield  {journal} {\bibinfo  {journal} {Phys. Rev. B}\ }\textbf {\bibinfo
  {volume} {90}},\ \bibinfo {pages} {125123} (\bibinfo {year}
  {2013})}\BibitemShut {NoStop}%
\bibitem [{\citenamefont {Yu}\ and\ \citenamefont {Zunger}(2014)}]{yu14}%
  \BibitemOpen
  \bibfield  {author} {\bibinfo {author} {\bibfnamefont {L.}~\bibnamefont
  {Yu}}\ and\ \bibinfo {author} {\bibfnamefont {A.}~\bibnamefont {Zunger}},\
  }\href@noop {} {\bibfield  {journal} {\bibinfo  {journal} {Nat. Commun.}\
  }\textbf {\bibinfo {volume} {5}},\ \bibinfo {pages} {5118} (\bibinfo {year}
  {2014})}\BibitemShut {NoStop}%
\bibitem [{\citenamefont {Lin}\ and\ \citenamefont {Demkov}(2014)}]{lin14}%
  \BibitemOpen
  \bibfield  {author} {\bibinfo {author} {\bibfnamefont {C.}~\bibnamefont
  {Lin}}\ and\ \bibinfo {author} {\bibfnamefont {A.~A.}\ \bibnamefont
  {Demkov}},\ }\href@noop {} {\bibfield  {journal} {\bibinfo  {journal} {Phys.
  Rev. Lett.}\ }\textbf {\bibinfo {volume} {113}},\ \bibinfo {pages} {157602}
  (\bibinfo {year} {2014})}\BibitemShut {NoStop}%
\bibitem [{\citenamefont {Pavlenko}\ \emph {et~al.}(2013)\citenamefont
  {Pavlenko}, \citenamefont {Kopp},\ and\ \citenamefont {Mannhart}}]{pav13}%
  \BibitemOpen
  \bibfield  {author} {\bibinfo {author} {\bibfnamefont {N.}~\bibnamefont
  {Pavlenko}}, \bibinfo {author} {\bibfnamefont {T.}~\bibnamefont {Kopp}}, \
  and\ \bibinfo {author} {\bibfnamefont {J.}~\bibnamefont {Mannhart}},\
  }\href@noop {} {\bibfield  {journal} {\bibinfo  {journal} {Phys. Rev. B}\
  }\textbf {\bibinfo {volume} {88}},\ \bibinfo {pages} {201104(R)} (\bibinfo
  {year} {2013})}\BibitemShut {NoStop}%
\bibitem [{\citenamefont {Zener}(1951)}]{zen51}%
  \BibitemOpen
  \bibfield  {author} {\bibinfo {author} {\bibfnamefont {C.}~\bibnamefont
  {Zener}},\ }\href@noop {} {\bibfield  {journal} {\bibinfo  {journal} {Phys.
  Rev.}\ }\textbf {\bibinfo {volume} {82}},\ \bibinfo {pages} {403} (\bibinfo
  {year} {1951})}\BibitemShut {NoStop}%
\bibitem [{\citenamefont {Anderson}\ and\ \citenamefont
  {Hasegawa}(1955)}]{and55}%
  \BibitemOpen
  \bibfield  {author} {\bibinfo {author} {\bibfnamefont {P.~W.}\ \bibnamefont
  {Anderson}}\ and\ \bibinfo {author} {\bibfnamefont {H.}~\bibnamefont
  {Hasegawa}},\ }\href@noop {} {\bibfield  {journal} {\bibinfo  {journal}
  {Phys. Rev.}\ }\textbf {\bibinfo {volume} {100}},\ \bibinfo {pages} {675}
  (\bibinfo {year} {1955})}\BibitemShut {NoStop}%
\bibitem [{\citenamefont {Hwang}\ \emph {et~al.}(2012)\citenamefont {Hwang},
  \citenamefont {Iwasa}, \citenamefont {Kawasaki}, \citenamefont {Keimer},
  \citenamefont {Nagaosa},\ and\ \citenamefont {Tokura}}]{hwa12}%
  \BibitemOpen
  \bibfield  {author} {\bibinfo {author} {\bibfnamefont {H.~Y.}\ \bibnamefont
  {Hwang}}, \bibinfo {author} {\bibfnamefont {Y.}~\bibnamefont {Iwasa}},
  \bibinfo {author} {\bibfnamefont {M.}~\bibnamefont {Kawasaki}}, \bibinfo
  {author} {\bibfnamefont {B.}~\bibnamefont {Keimer}}, \bibinfo {author}
  {\bibfnamefont {N.}~\bibnamefont {Nagaosa}}, \ and\ \bibinfo {author}
  {\bibfnamefont {Y.}~\bibnamefont {Tokura}},\ }\href@noop {} {\bibfield
  {journal} {\bibinfo  {journal} {Nature Materials}\ }\textbf {\bibinfo
  {volume} {11}},\ \bibinfo {pages} {103} (\bibinfo {year} {2012})}\BibitemShut
  {NoStop}%
\bibitem [{\citenamefont {Rubtsov}\ \emph {et~al.}(2005)\citenamefont
  {Rubtsov}, \citenamefont {Savkin},\ and\ \citenamefont
  {Lichtenstein}}]{rub05}%
  \BibitemOpen
  \bibfield  {author} {\bibinfo {author} {\bibfnamefont {A.~N.}\ \bibnamefont
  {Rubtsov}}, \bibinfo {author} {\bibfnamefont {V.~V.}\ \bibnamefont {Savkin}},
  \ and\ \bibinfo {author} {\bibfnamefont {A.~I.}\ \bibnamefont
  {Lichtenstein}},\ }\href@noop {} {\bibfield  {journal} {\bibinfo  {journal}
  {Phys. Rev. B}\ }\textbf {\bibinfo {volume} {72}},\ \bibinfo {pages} {035122}
  (\bibinfo {year} {2005})}\BibitemShut {NoStop}%
\bibitem [{\citenamefont {Werner}\ \emph {et~al.}(2006)\citenamefont {Werner},
  \citenamefont {Comanac}, \citenamefont {de' Medici}, \citenamefont {Troyer},\
  and\ \citenamefont {Millis}}]{wer06}%
  \BibitemOpen
  \bibfield  {author} {\bibinfo {author} {\bibfnamefont {P.}~\bibnamefont
  {Werner}}, \bibinfo {author} {\bibfnamefont {A.}~\bibnamefont {Comanac}},
  \bibinfo {author} {\bibfnamefont {L.}~\bibnamefont {de' Medici}}, \bibinfo
  {author} {\bibfnamefont {M.}~\bibnamefont {Troyer}}, \ and\ \bibinfo {author}
  {\bibfnamefont {A.~J.}\ \bibnamefont {Millis}},\ }\href@noop {} {\bibfield
  {journal} {\bibinfo  {journal} {Phys. Rev. Lett.}\ }\textbf {\bibinfo
  {volume} {97}},\ \bibinfo {pages} {076405} (\bibinfo {year}
  {2006})}\BibitemShut {NoStop}%
\bibitem [{\citenamefont {Parcollet}\ \emph {et~al.}(2015)\citenamefont
  {Parcollet}, \citenamefont {Ferrero}, \citenamefont {Ayral}, \citenamefont
  {Hafermann}, \citenamefont {Krivenko}, \citenamefont {Messio},\ and\
  \citenamefont {Seth}}]{triqs_code}%
  \BibitemOpen
  \bibfield  {author} {\bibinfo {author} {\bibfnamefont {O.}~\bibnamefont
  {Parcollet}}, \bibinfo {author} {\bibfnamefont {M.}~\bibnamefont {Ferrero}},
  \bibinfo {author} {\bibfnamefont {T.}~\bibnamefont {Ayral}}, \bibinfo
  {author} {\bibfnamefont {H.}~\bibnamefont {Hafermann}}, \bibinfo {author}
  {\bibfnamefont {I.}~\bibnamefont {Krivenko}}, \bibinfo {author}
  {\bibfnamefont {L.}~\bibnamefont {Messio}}, \ and\ \bibinfo {author}
  {\bibfnamefont {P.}~\bibnamefont {Seth}},\ }\href@noop {} {\bibfield
  {journal} {\bibinfo  {journal} {Computer Physics Communications}\ }\textbf
  {\bibinfo {volume} {196}},\ \bibinfo {pages} {398} (\bibinfo {year}
  {2015})}\BibitemShut {NoStop}%
\bibitem [{\citenamefont {Boehnke}\ \emph {et~al.}(2011)\citenamefont
  {Boehnke}, \citenamefont {Hafermann}, \citenamefont {Ferrero}, \citenamefont
  {Lechermann},\ and\ \citenamefont {Parcollet}}]{boe11}%
  \BibitemOpen
  \bibfield  {author} {\bibinfo {author} {\bibfnamefont {L.}~\bibnamefont
  {Boehnke}}, \bibinfo {author} {\bibfnamefont {H.}~\bibnamefont {Hafermann}},
  \bibinfo {author} {\bibfnamefont {M.}~\bibnamefont {Ferrero}}, \bibinfo
  {author} {\bibfnamefont {F.}~\bibnamefont {Lechermann}}, \ and\ \bibinfo
  {author} {\bibfnamefont {O.}~\bibnamefont {Parcollet}},\ }\href@noop {}
  {\bibfield  {journal} {\bibinfo  {journal} {Phys. Rev. B}\ }\textbf {\bibinfo
  {volume} {84}},\ \bibinfo {pages} {075145} (\bibinfo {year}
  {2011})}\BibitemShut {NoStop}%
\bibitem [{\citenamefont {Lechermann}\ \emph {et~al.}(2007)\citenamefont
  {Lechermann}, \citenamefont {Georges}, \citenamefont {Kotliar},\ and\
  \citenamefont {Parcollet}}]{lec07}%
  \BibitemOpen
  \bibfield  {author} {\bibinfo {author} {\bibfnamefont {F.}~\bibnamefont
  {Lechermann}}, \bibinfo {author} {\bibfnamefont {A.}~\bibnamefont {Georges}},
  \bibinfo {author} {\bibfnamefont {G.}~\bibnamefont {Kotliar}}, \ and\
  \bibinfo {author} {\bibfnamefont {O.}~\bibnamefont {Parcollet}},\ }\href@noop
  {} {\bibfield  {journal} {\bibinfo  {journal} {Phys. Rev. B}\ }\textbf
  {\bibinfo {volume} {76}},\ \bibinfo {pages} {155102} (\bibinfo {year}
  {2007})}\BibitemShut {NoStop}%
\bibitem [{\citenamefont {Amadon}\ \emph {et~al.}(2008)\citenamefont {Amadon},
  \citenamefont {Lechermann}, \citenamefont {Georges}, \citenamefont {Jollet},
  \citenamefont {Wehling},\ and\ \citenamefont {Lichtenstein}}]{ama08}%
  \BibitemOpen
  \bibfield  {author} {\bibinfo {author} {\bibfnamefont {B.}~\bibnamefont
  {Amadon}}, \bibinfo {author} {\bibfnamefont {F.}~\bibnamefont {Lechermann}},
  \bibinfo {author} {\bibfnamefont {A.}~\bibnamefont {Georges}}, \bibinfo
  {author} {\bibfnamefont {F.}~\bibnamefont {Jollet}}, \bibinfo {author}
  {\bibfnamefont {T.~O.}\ \bibnamefont {Wehling}}, \ and\ \bibinfo {author}
  {\bibfnamefont {A.~I.}\ \bibnamefont {Lichtenstein}},\ }\href@noop {}
  {\bibfield  {journal} {\bibinfo  {journal} {Phys. Rev. B}\ }\textbf {\bibinfo
  {volume} {77}},\ \bibinfo {pages} {205112} (\bibinfo {year}
  {2008})}\BibitemShut {NoStop}%
\bibitem [{\citenamefont {Li}\ \emph {et~al.}(1989)\citenamefont {Li},
  \citenamefont {W\"olfle},\ and\ \citenamefont {Hirschfeld}}]{li89}%
  \BibitemOpen
  \bibfield  {author} {\bibinfo {author} {\bibfnamefont {T.}~\bibnamefont
  {Li}}, \bibinfo {author} {\bibfnamefont {P.}~\bibnamefont {W\"olfle}}, \ and\
  \bibinfo {author} {\bibfnamefont {P.~J.}\ \bibnamefont {Hirschfeld}},\
  }\href@noop {} {\bibfield  {journal} {\bibinfo  {journal} {Phys. Rev. B}\
  }\textbf {\bibinfo {volume} {40}},\ \bibinfo {pages} {6817} (\bibinfo {year}
  {1989})}\BibitemShut {NoStop}%
\bibitem [{\citenamefont {B\"{u}nemann}\ \emph {et~al.}(1998)\citenamefont
  {B\"{u}nemann}, \citenamefont {Weber},\ and\ \citenamefont
  {Gebhard}}]{bue98}%
  \BibitemOpen
  \bibfield  {author} {\bibinfo {author} {\bibfnamefont {J.}~\bibnamefont
  {B\"{u}nemann}}, \bibinfo {author} {\bibfnamefont {W.}~\bibnamefont {Weber}},
  \ and\ \bibinfo {author} {\bibfnamefont {F.}~\bibnamefont {Gebhard}},\
  }\href@noop {} {\bibfield  {journal} {\bibinfo  {journal} {Phys. Rev. B}\
  }\textbf {\bibinfo {volume} {57}},\ \bibinfo {pages} {6896} (\bibinfo {year}
  {1998})}\BibitemShut {NoStop}%
\bibitem [{\citenamefont {Lechermann}(2009)}]{lec09}%
  \BibitemOpen
  \bibfield  {author} {\bibinfo {author} {\bibfnamefont {F.}~\bibnamefont
  {Lechermann}},\ }\href@noop {} {\bibfield  {journal} {\bibinfo  {journal}
  {Phys. Rev. Lett.}\ }\textbf {\bibinfo {volume} {102}},\ \bibinfo {pages}
  {046403} (\bibinfo {year} {2009})}\BibitemShut {NoStop}%
\bibitem [{\citenamefont {Huang}\ \emph {et~al.}(2012)\citenamefont {Huang},
  \citenamefont {Du},\ and\ \citenamefont {Dai}}]{hua12}%
  \BibitemOpen
  \bibfield  {author} {\bibinfo {author} {\bibfnamefont {L.}~\bibnamefont
  {Huang}}, \bibinfo {author} {\bibfnamefont {L.}~\bibnamefont {Du}}, \ and\
  \bibinfo {author} {\bibfnamefont {X.}~\bibnamefont {Dai}},\ }\href@noop {}
  {\bibfield  {journal} {\bibinfo  {journal} {Phys. Rev. B}\ }\textbf {\bibinfo
  {volume} {86}},\ \bibinfo {pages} {035150} (\bibinfo {year}
  {2012})}\BibitemShut {NoStop}%
\bibitem [{\citenamefont {Andrade}\ \emph {et~al.}(2009)\citenamefont
  {Andrade}, \citenamefont {Miranda},\ and\ \citenamefont
  {Dobrosavljevi{\'c}}}]{and09}%
  \BibitemOpen
  \bibfield  {author} {\bibinfo {author} {\bibfnamefont {E.~C.}\ \bibnamefont
  {Andrade}}, \bibinfo {author} {\bibfnamefont {E.}~\bibnamefont {Miranda}}, \
  and\ \bibinfo {author} {\bibfnamefont {V.}~\bibnamefont
  {Dobrosavljevi{\'c}}},\ }\href@noop {} {\bibfield  {journal} {\bibinfo
  {journal} {Phys. Rev. Lett.}\ }\textbf {\bibinfo {volume} {102}},\ \bibinfo
  {pages} {206403} (\bibinfo {year} {2009})}\BibitemShut {NoStop}%
\bibitem [{\citenamefont {Kotliar}\ and\ \citenamefont
  {Ruckenstein}(1986)}]{kot86}%
  \BibitemOpen
  \bibfield  {author} {\bibinfo {author} {\bibfnamefont {G.}~\bibnamefont
  {Kotliar}}\ and\ \bibinfo {author} {\bibfnamefont {A.~E.}\ \bibnamefont
  {Ruckenstein}},\ }\href@noop {} {\bibfield  {journal} {\bibinfo  {journal}
  {Phys. Rev. Lett.}\ }\textbf {\bibinfo {volume} {57}},\ \bibinfo {pages}
  {1362} (\bibinfo {year} {1986})}\BibitemShut {NoStop}%
\bibitem [{\citenamefont {Lechermann}\ \emph {et~al.}(2005)\citenamefont
  {Lechermann}, \citenamefont {Biermann},\ and\ \citenamefont
  {Georges}}]{lecproc}%
  \BibitemOpen
  \bibfield  {author} {\bibinfo {author} {\bibfnamefont {F.}~\bibnamefont
  {Lechermann}}, \bibinfo {author} {\bibfnamefont {S.}~\bibnamefont
  {Biermann}}, \ and\ \bibinfo {author} {\bibfnamefont {A.}~\bibnamefont
  {Georges}},\ }\href@noop {} {\bibfield  {journal} {\bibinfo  {journal} {Prog.
  Theor. Phys. Suppl.}\ }\textbf {\bibinfo {volume} {160}},\ \bibinfo {pages}
  {233} (\bibinfo {year} {2005})}\BibitemShut {NoStop}%
\bibitem [{\citenamefont {Fitzsimmons}\ \emph {et~al.}(2011)\citenamefont
  {Fitzsimmons}, \citenamefont {Hengartner}, \citenamefont {Singh},
  \citenamefont {Zhernenkov}, \citenamefont {Bruno}, \citenamefont
  {Santamaria}, \citenamefont {Brinkman}, \citenamefont {Huijben},
  \citenamefont {Molegraaf}, \citenamefont {de~la Venta}, ,\ and\ \citenamefont
  {Schuller}}]{fit11}%
  \BibitemOpen
  \bibfield  {author} {\bibinfo {author} {\bibfnamefont {M.~R.}\ \bibnamefont
  {Fitzsimmons}}, \bibinfo {author} {\bibfnamefont {N.~W.}\ \bibnamefont
  {Hengartner}}, \bibinfo {author} {\bibfnamefont {S.}~\bibnamefont {Singh}},
  \bibinfo {author} {\bibfnamefont {M.}~\bibnamefont {Zhernenkov}}, \bibinfo
  {author} {\bibfnamefont {F.~Y.}\ \bibnamefont {Bruno}}, \bibinfo {author}
  {\bibfnamefont {J.}~\bibnamefont {Santamaria}}, \bibinfo {author}
  {\bibfnamefont {A.}~\bibnamefont {Brinkman}}, \bibinfo {author}
  {\bibfnamefont {M.}~\bibnamefont {Huijben}}, \bibinfo {author} {\bibfnamefont
  {H.~J.~A.}\ \bibnamefont {Molegraaf}}, \bibinfo {author} {\bibfnamefont
  {J.}~\bibnamefont {de~la Venta}}, , \ and\ \bibinfo {author} {\bibfnamefont
  {I.~K.}\ \bibnamefont {Schuller}},\ }\href@noop {} {\bibfield  {journal}
  {\bibinfo  {journal} {Phys. Rev. Lett.}\ }\textbf {\bibinfo {volume} {107}},\
  \bibinfo {pages} {217201} (\bibinfo {year} {2011})}\BibitemShut {NoStop}%
\bibitem [{\citenamefont {Ron}\ \emph {et~al.}(2014)\citenamefont {Ron},
  \citenamefont {Maniv}, \citenamefont {Graf}, \citenamefont {Park},\ and\
  \citenamefont {Dagan}}]{ron14}%
  \BibitemOpen
  \bibfield  {author} {\bibinfo {author} {\bibfnamefont {A.}~\bibnamefont
  {Ron}}, \bibinfo {author} {\bibfnamefont {E.}~\bibnamefont {Maniv}}, \bibinfo
  {author} {\bibfnamefont {D.}~\bibnamefont {Graf}}, \bibinfo {author}
  {\bibfnamefont {J.-H.}\ \bibnamefont {Park}}, \ and\ \bibinfo {author}
  {\bibfnamefont {Y.}~\bibnamefont {Dagan}},\ }\href@noop {} {\bibfield
  {journal} {\bibinfo  {journal} {Phys. Rev. Lett.}\ }\textbf {\bibinfo
  {volume} {113}},\ \bibinfo {pages} {216801} (\bibinfo {year}
  {2014})}\BibitemShut {NoStop}%
\bibitem [{\citenamefont {Wang}\ \emph {et~al.}(2015)\citenamefont {Wang},
  \citenamefont {Pan}, \citenamefont {Song}, \citenamefont {Mi}, \citenamefont
  {Zou}, \citenamefont {Wang},\ and\ \citenamefont {Zhang}}]{wan15}%
  \BibitemOpen
  \bibfield  {author} {\bibinfo {author} {\bibfnamefont {S.}~\bibnamefont
  {Wang}}, \bibinfo {author} {\bibfnamefont {L.}~\bibnamefont {Pan}}, \bibinfo
  {author} {\bibfnamefont {J.-J.}\ \bibnamefont {Song}}, \bibinfo {author}
  {\bibfnamefont {W.}~\bibnamefont {Mi}}, \bibinfo {author} {\bibfnamefont
  {J.-J.}\ \bibnamefont {Zou}}, \bibinfo {author} {\bibfnamefont
  {L.}~\bibnamefont {Wang}}, \ and\ \bibinfo {author} {\bibfnamefont
  {X.}~\bibnamefont {Zhang}},\ }\href@noop {} {\bibfield  {journal} {\bibinfo
  {journal} {J. Am. Chem. Soc.}\ } (\bibinfo {year} {2015})}\BibitemShut
  {NoStop}%
\end{thebibliography}%

\end{document}